\documentclass[sigconf,screen]{acmart}

\AtBeginDocument{%
  \providecommand\BibTeX{{%
    \normalfont B\kern-0.5em{\scshape i\kern-0.25em b}\kern-0.8em\TeX}}}

\copyrightyear{2025}
\acmYear{2025}
\setcopyright{rightsretained}
\acmConference[CHI '25]{CHI Conference on Human Factors in Computing Systems}{April 26-May 1, 2025}{Yokohama, Japan}
\acmBooktitle{CHI Conference on Human Factors in Computing Systems (CHI '25), April 26-May 1, 2025, Yokohama, Japan}\acmDOI{10.1145/3706598.3713379}
\acmISBN{979-8-4007-1394-1/25/04}

\usepackage{xcolor}
\definecolor{pumpkin}{RGB}{211, 84, 0}

\usepackage[utf8]{inputenc}
\usepackage{CJKutf8}
\usepackage{soul}
\usepackage{multirow}
\usepackage{float}
\usepackage{graphicx}
\usepackage{longtable}
\usepackage{makecell}
\usepackage{booktabs}
\usepackage{arydshln}
\usepackage{tabularx}
\usepackage{stfloats}
\usepackage{enumitem}
\usepackage{cellspace}
\usepackage{subcaption}

\begin{document}
\begin{CJK*}{UTF8}{gbsn}

\title[Hashtag Re-Appropriation for Audience Control]{Hashtag Re-Appropriation for Audience Control on Recommendation-Driven Social Media Xiaohongshu (rednote) }

\author{Ruyuan Wan}
\authornote{Both authors contributed equally to this research.}
\affiliation{%
  \institution{The Pennsylvania State University}
  \city{University Park}
  \state{PA}
  \country{USA}
}
\email{rjw6289@psu.edu}

\author{Lingbo Tong}
\authornotemark[1]
\affiliation{%
  \institution{University of Notre Dame}
  \city{Notre Dame}
  \state{IN}
  \country{USA}
}
\email{ltong2@nd.edu}

\author{Tiffany Knearem}
\affiliation{%
  \institution{Google}
  \city{Cambridge}
  \state{MA}
  \country{USA}}
\email{tknearem@gmail.com }

\author{Toby Jia-Jun Li}
\affiliation{%
  \institution{University of Notre Dame}
  \city{Notre Dame}
  \state{IN}
  \country{USA}}
\email{toby.j.li@nd.edu}

\author{Ting-Hao `Kenneth' Huang }
\affiliation{%
  \institution{The Pennsylvania State University}
  \city{University Park}
  \state{PA}
  \country{USA}}
\email{txh710@psu.edu}

\author{Qunfang Wu}
\affiliation{%
  \institution{Harvard University}
  \city{Cambridge}
  \state{MA}
  \country{USA}}
\email{qunfangwu@fas.harvard.edu}

\renewcommand{\shortauthors}{Wan and Tong, et al.}

\begin{abstract}
Algorithms have played a central role in personalized recommendations on social media. However, they also present significant obstacles for content creators trying to predict and manage their audience reach. This issue is particularly challenging for marginalized groups seeking to maintain safe spaces. Our study explores how women on Xiaohongshu (rednote), a recommendation-driven social platform, proactively re-appropriate hashtags (e.g., \#宝宝辅食, Baby Supplemental Food) by using them in posts unrelated to their literal meaning. The hashtags were strategically chosen from topics that would be uninteresting to the male audience they wanted to block.
Through a mixed-methods approach, we analyzed the practice of hashtag re-appropriation based on 5,800 collected posts and interviewed 24 active users from diverse backgrounds to uncover users' motivations and reactions towards the re-appropriation. This practice highlights how users can reclaim agency over content distribution on recommendation-driven platforms, offering insights into self-governance within algorithmic-centered power structures.
\end{abstract}

\ccsdesc[500]{Human-centered computing~Empirical studies in collaborative and social computing}

\keywords{Social Media Governance, Content Moderation, Feminist HCI, Hashtag Activism, Platform Design}

\maketitle

\section{Introduction}
\everypar{\looseness=-1}
The rise in popularity of recommendation-driven social media plays a pivotal role in modern communication by offering users avenues of connection and information exchange ~\cite{guy2010social, eirinaki2018recommender, song2019session, nilashi2021recommendation}. Currently, eight of the ten most popular global social media platforms are powered by recommendation algorithms---namely, Facebook, YouTube, Instagram, TikTok, Wechat, Telegram, Snapchat, and Douyin---with a combined total of 15.96 billion monthly active users worldwide ~\cite{statista_global_social}. 
Equipped with recommendation algorithms, these platforms analyze user behavior and preferences to provide personalized content feeds~\cite{qian2013personalized, khater2017personalized, vombatkere2024tiktok}, which helps users discover unlimited new information and keep them engaged. 

However, these platforms also introduce challenges due to the opaque nature of recommendation algorithms. The lack of transparency in these algorithms diminishes users' agency over what they see and who sees their posts ~\cite{Devito2017affordance}. While some platforms provide features such as editing topic preferences to adjust recommendation feeds and users employ content moderation strategies like using hashtags to influence audience reach, existing studies highlight users still struggle with issues like echo chambers, low visibility, and online harassment ~\cite{steen2023you, thomas2022s, gao2023echo}.


Moreover, the opaque nature of recommendation mechanisms presents unique challenges for groups such as women, Black people, and LGBTQ+ individuals~\cite{klassen2021more, karizat2021algorithmic}. 
\citet{karizat2021algorithmic} found that TikTok's ``For You'' Page algorithm actively suppresses content related to marginalized social identities. 
Experienced the restriction of content visibility on TikTok, content creators increasingly invent words to bypass content moderation algorithms from banning their videos, such as using ``le\$bian'' for ``lesbian'' ~\cite{steen2023you}.
Also, the lack of direct control over content visibility could result in ``context collapse,'' when content intended for a single audience is exposed to diverse audiences within one environment, leading to unexpected or even negative consequences for their self-presentation~\cite{context-collapse-definition, marwick2011tweet}. For example, on Black Twitter, context collapse occurs when tweets are accidentally exposed to audiences outside of the Black community, making it difficult for users to manage self-presentation and avoid miscommunication ~\cite{klassen2021more, black-twitter-woke}. Given algorithms' opaqueness, users have limited ways to predict or control their audience~\cite{devito2018people,Devito2021adaptive,devito2017algorithms,Devito2017affordance}. 
Content creators can add hashtags to extend their content, while content consumers can use hashtags as keywords to search content; recommendation algorithms leverage hashtags to deliver posts to targeted audiences ~\cite{devito2018people, TikTokSupport}, further shaping the visibility dynamics. Recognizing the capability of hashtags, users have developed increasingly creative approaches to optimize their use. 

In this paper, we investigated the widespread use of the hashtag \#\begin{CJK*}{UTF8}{gbsn}宝宝辅食\end{CJK*} (Baby Supplemental Food, abbreviated as BSF)\footnote{The quote shows the original Chinese hashtags and its English translation and abbreviation in the parentheses.} on the Chinese social media Xiaohongshu\footnote{Xiaohongshu means ``Little Red Book,'' also known as rednote in English. It is a social sharing and e-commerce platform. As of 2024, Xiaohongshu has over 209 million monthly active users ~\cite{MAU_xiaohongshu}. \url{https://www.xiaohongshu.com/}}. 
\#BSF became viral in the Xiaohongshu community for its unexpected and diverse usage, which has been widely reported and discussed online ~\cite{sohu2024, Pengpai2024, xiaohongshu2024} \footnote{These articles are: 
{\begin{CJK*}{UTF8}{gbsn}宝宝辅食与小红书隔“男”墙\end{CJK*} (Baby Supplemental Food and Xiaohongshu's wall against ``men'')} on Sohu,  {``\begin{CJK*}{UTF8}{gbsn}宝宝辅食''tag：互联网新型护盾\end{CJK*} (``Baby Supplemental Food'' tag: a new type of internet shield)} on Pengpai, 
and {\begin{CJK*}{UTF8}{gbsn}新传热点宝宝辅食?更适合躲避男凝的标签\end{CJK*} (News communication hot topic Baby Supplemental Food? a better tag to avoid the male gaze)} on Xiaohongshu.}. 
The articles reported many Xiaohongshu female users re-appropriate \#BSF in an attempt to moderate the recommendation algorithm not push their posts to male users, instead of tagging baby food-related posts. The hashtag re-appropriation practice reflects women's effort to manage their safe spaces under the challenges of recommendation-driven platforms like Xiaohongshu. Driven by this initial observation, we systematically investigate the phenomenon from a mixed-methods approach. 
Unlike network-based platforms such as Facebook and Instagram, where content delivery primarily depends on user social networks, Xiaohongshu delivers content across the entire platform based on user behavior and interests~\cite{zhang2024form}, making audiences beyond their direct followers or group memberships~\cite{cao2024voices, johnson2022s}.
On the other hand, in contrast to community-based platforms like Reddit---where user engagement is governed by community rules and moderators, behaviors on Xiaohongshu are regulated mainly by general platform rules and moderation~\cite{xiaohongshu_guidelines}. These differences raise the question of how users can gain more control over algorithms---both in terms of what they see and who sees their posts---and maintain safe spaces on recommendation-driven platforms. Building on prior literature on hashtag activism~\cite{mueller2021demographic, yang2016narrative}, feminism~\cite{d2023data, bardzell2010feminist}, and user control in algorithmic platforms~\cite{witzenberger2018hyperdodge,van2018networks,burrell_when_2019}, our study aims to address the following research questions:

\begin{itemize}
    \item \textbf{RQ1}: What were the hashtag usage and evolution patterns in the \#BSF hashtag appropriation?
    
    \item \textbf{RQ2:} What were the motivations, attitudes, and reactions of different user groups around this hashtag re-appropriation practice? 
\end{itemize}

We conducted a mixed-methods study to investigate the \#BSF case on Xiaohongshu. We collected posts with \#BSF and related hashtags on Xiaohongshu via web scraping and performed four steps of quantitative analysis: hashtag-post relevance analysis, topic modeling, post expression analysis, and hashtag co-occurrence network analysis, to reveal hashtag usage patterns (RQ1). Additionally, we interviewed 24 active Xiaohongshu users (17 women and seven men) to explore the evolution of hashtag use, and users' motivations and reactions (RQ2).

Our research reveals the practice of \#BSF hashtag re-appropriation as a process of blocking, attracting, and evolving. Due to ineffective reporting and blocking mechanisms and a loss of trust in the platform’s women-user-friendly stance, women began to adopt hashtag re-appropriation as a strategy. Different user groups responded to this re-appropriation with varying degrees of support and misunderstanding. The case of hashtag re-appropriation enabled women to shape their narratives through interconnected posts while simultaneously blocking unwanted viewers, demonstrating a form of everyday resistance to digital feminism in China. It also inspires the broader implications of audience control on recommendation-driven platforms and the potential for user self-governance within algorithm-centered structures.

\section{Related Work}
\everypar{\looseness=-1}

\subsection{Digital Feminism Practices in China}

Digital feminism is rooted in the fourth wave of feminism, which has fostered a ``call-out'' culture to confront sexist behaviors ~\cite{munro2013feminism}. Social media has become crucial in this movement, advancing idea sharing and a broader understanding of experienced oppression  ~\cite{evans2015critical}. 
The worldwide \#MeToo movement has encouraged victims of sexual violence to speak out about their experiences~\cite{mendes2018metoo}. In 2018, the movement reached Chinese universities, where students publicized their experiences of being harassed by their instructors online ~\cite{lin2019individual}. Since then, the movement has spread across China, with victims increasingly using social media to share their stories and discuss gender-related issues, such as discrimination, harassment, and violence~\cite{tangshan, huang2021metoo, wang2023counter}.

Chinese digital feminism is primarily driven by young, urban women with college education ~\cite{hou2020rewriting, yin2021intersectional}. Their mothers' generation experienced the state-led gender affirmative action movement, with popular slogans like ``men and women are the same'' and ``women can hold up half the sky''~\cite{rural-feminism}.
Additionally, the one-child policy generation of urban Chinese women, born in the 1980s and 1990s, benefited from greater family resources and less overt gender bias in their early childhood~\cite{c-fem, xie2021women}. However, subsequent experiences of inequality in education and work cultivated their strong gender consciousness and aversion to gender disparities~\cite{c-fem}.


As digital feminism develops, anti-feminists have also become active on Chinese social media. They use the derogatory term ``\begin{CJK*}{UTF8}{gbsn}田园女权\end{CJK*} (Rural Feminism)'' to attack women users~\cite{rural-feminism, c-fem}, creating a false dichotomy between ``rural'' and ``authentic'' feminism. They condemn certain feminist demands (such as men taking on more household responsibilities) while endorsing ``moderate'' claims (such as women rejecting bride prices)~\cite{rural-feminism}.
Opposition to feminism also exists among women, who criticize it as a hidden disciplinary force that undermines female solidarity (such as wearing makeup)~\cite{women-hate-feminists}. Some also view digital feminism as merely a trendy business~\cite{women-hate-feminists}.
Additionally, a large nuanced subgroup exists distinct from feminists or anti-feminists. ~\citet{yin2021intersectional} used the term ``pro-change counterpublics'' to describe those who refuse to identify as feminists but actively protest against sexual harassment in online discussions. This subgroup differs from those who fully embrace feminist ideas. 

As a result of divisions in Chinese digital feminism, inclusivity has become a significant challenge. 
Moreover, the increasingly restricted social media environment poses additional challenges for digital feminism to maintain its independence while navigating mainstream ideologies and platform regulations~\cite{yaya, grassroots, pink}. Women face a unique ecosystem in the Chinese online environment given historical and cultural factors. By examining the practices of women in Xiaohongshu, our study provides a new perspective bridging localized dynamics with broader global issues in digital feminism practice.



\subsection{Feminism Theory in HCI Practices}
The evolution of feminist discourse has profoundly shaped our understanding of how gender is conceptualized within social systems.
Beginning with the second wave of feminism, Simone de Beauvoir's assertion, ``One is not born a woman, but becomes one''~\cite{de1981one} laid the groundwork for examining how gender is constructed through social processes. Then the third wave of feminism expanded the focus to explore the complexities of gender identity. Further, feminist standpoint theory emphasizes the value of women's unique experiences, challenging traditional social science paradigms dominated by patriarchal assumptions~\cite{harding2004feminist}. 

As feminist scholarship evolved, critiques of White liberal feminism led to the development of Black feminism, postcolonial feminism, and intersectionality. ~\citet{crenshaw2013demarginalizing} introduced intersectionality to address how traditional feminist theory often overlooked the complex experiences of Black women,  considering multiple identity dimensions---such as race, sexual orientation, and nationality. 
Similarly, ~\citet{mohanty2003under} advanced a postcolonial feminist perspective that foregrounded ``feminist solidarity,'' highlighting the global connections among marginalized women through colonialism and mutuality in lived experiences to bridge differences across contexts.

Building on these theories, ~\citet{bardzell2010feminist} introduced a critical feminist agenda in HCI, advocating for gender-related design activities to embrace pluralism and advocacy. 
Bardzell argued that integrating feminist qualities into design could critique existing designs and identify opportunities to better meet women's needs. ~\citet{bardzell2012critical} also emphasized the principle of provocativeness in critical design, which balances a ``slight strangeness''---neither too strange to be dismissed nor too normal to be overlooked~\cite{dunne2001design}. However, achieving this balance proved challenging; even experienced design teams struggled to create artifacts that resonated with users as genuinely provocative, often only realizing the limitations of their designs after development and field tests. The women developers of Archive of Our Own (AO3) consciously incorporated a tagging feature into the website design, leading to a successful, women-dominated fan fiction archive. The case of AO3 exemplifies how feminist HCI design can successfully integrate community values and needs ~\cite{fiesler2016archive}. In contrast, a failed case is ~\citet{bardzell2010feminist}'s example of the “World Washer,” which neglected the fragility of South Indian women’s saris. As ~\citet{fiesler2016archive} highlighted, socio-technical systems design should prioritize user participation, address specific community needs, and avoid harmful universal solutions. ~\citet{sultana2018design} further suggested design guidelines within patriarchal society---supporting the tactics that women already used to cope with rather than focusing on the problems nested with systematic challenges and power dynamics. 

By incorporating diverse viewpoints, feminist scholars sought to avoid one-sided interpretations and foster inclusive technologies. Influenced by their intersectionality, women users' needs are sensitive and complex, which can be better understood through the theories. Informed by feminism theories both inside and outside of HCI, we examined women users'  hashtag re-appropriation practices on Xiaohongshu, considering their unique experiences. 


\subsection{Building Safety in Online Communities}

HCI scholarship has explored how community members collectively build safety, focusing on aspects like identity~\cite{dym2019coming,dosono2018identity,dosono2019moderation,massanari2017gamergate}, misinformation~\cite{starbird2011voluntweeters,huang2015connected}, and social rights~\cite{starbird2012will,wulf2013ground,stewart2017drawing} through discourse framing or technical solutions. Group members collectively construct a shared identity through sense-making and norm enforcement~\cite{dym2019coming,dosono2018identity,dosono2020decolonizing}. According to social identity theory, social groups are understood through individuals who define their self-concept by the characteristics of their shared social identity~\cite{riek2006intergroup}. 
For example, during the Black Lives Matter movement, hashtags served as gatekeepers, controlling access to the movement~\cite{stewart2017drawing}. ~\citet{klassen2021more} draw a parallel between Black Twitter and the Negro Motorist Green Book: both serve as tools for navigating systemic racism while fostering community, safety and empowerment for the Black community. Like the Green Book’s network of safe spaces, Black Twitter offers a space for collective identity formation, activism, and extends beyond the book providing broader real-time interaction.
Taken together, these works outline collective strategies that contributed to safer online spaces.

Women users also tried to build safe spaces through hashtags or private group chats on social media like Twitter and Facebook. ~\citet{golbeck2017hashtags} found that hashtags like \#distractinglySexy and \#iLookLikeAnEngineer not only raised awareness but also provided social support for participants in response to sexism in the sciences, fostering a sense of collective identity. Their study underscores the role of hashtags in forming ad hoc online communities. 
\citet{dixon2014feminist}'s research also shows hashtag feminism has created virtual spaces for those facing inequality to share their narratives and seek solidarity using hashtags like \#bringbackourgirls and \#YesAllWomen. Nevertheless, the study also highlights the challenge of harassment and hate speech arising in such highly visible and politicized environments. Recognizing the strategies and challenges in managing online safe spaces, our study explored how Chinese women users strategically construct safe spaces for their empowerment on Xiaohongshu. 

\begin{figure}[t]
    \centering
    \begin{minipage}[t]{0.5\linewidth}
        \centering
        \includegraphics[width=0.95\textwidth]{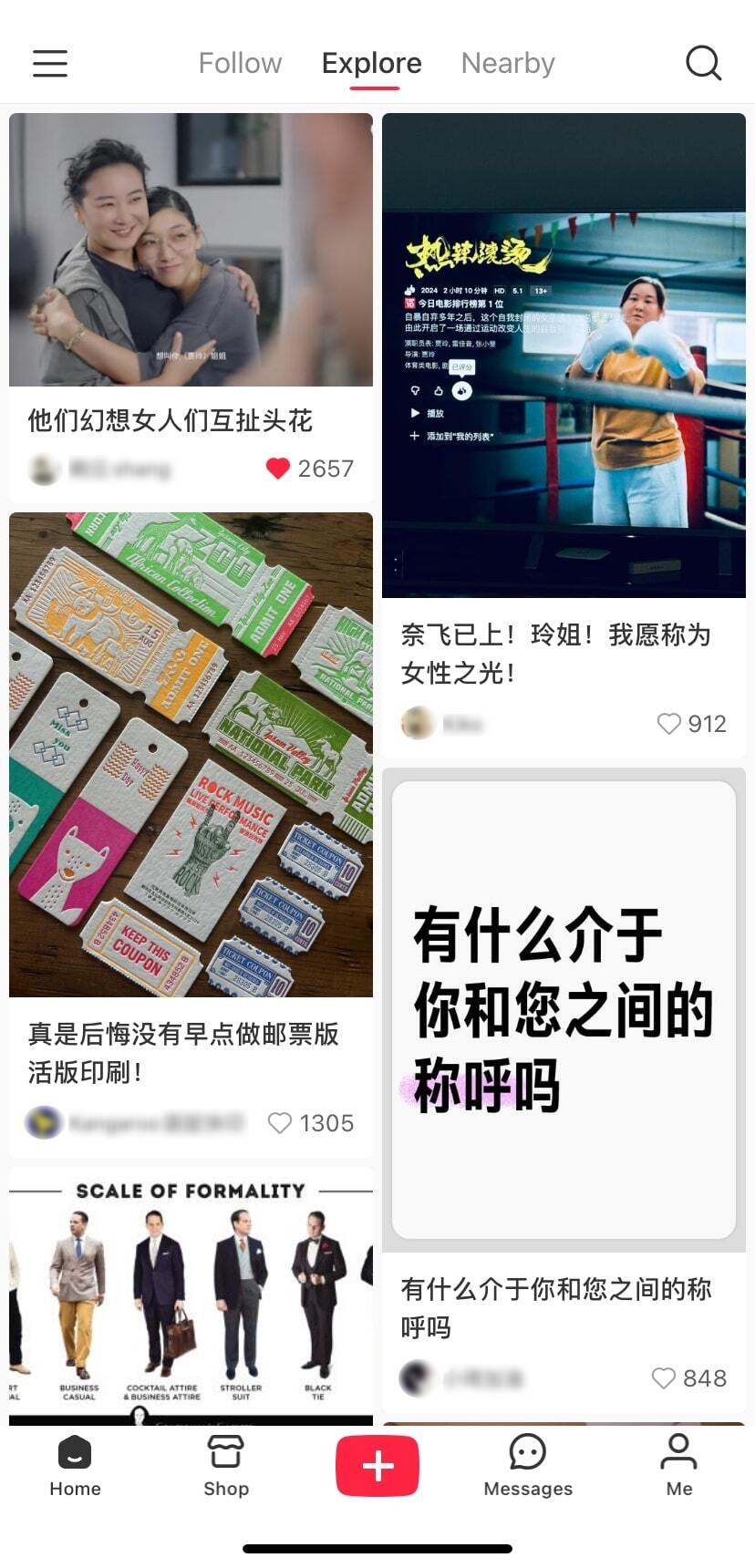}
        \caption*{(1) Explore page}
    \end{minipage}%
    \hfill
    \begin{minipage}[t]{0.5\linewidth}
        \centering
        \includegraphics[width=0.95\textwidth]{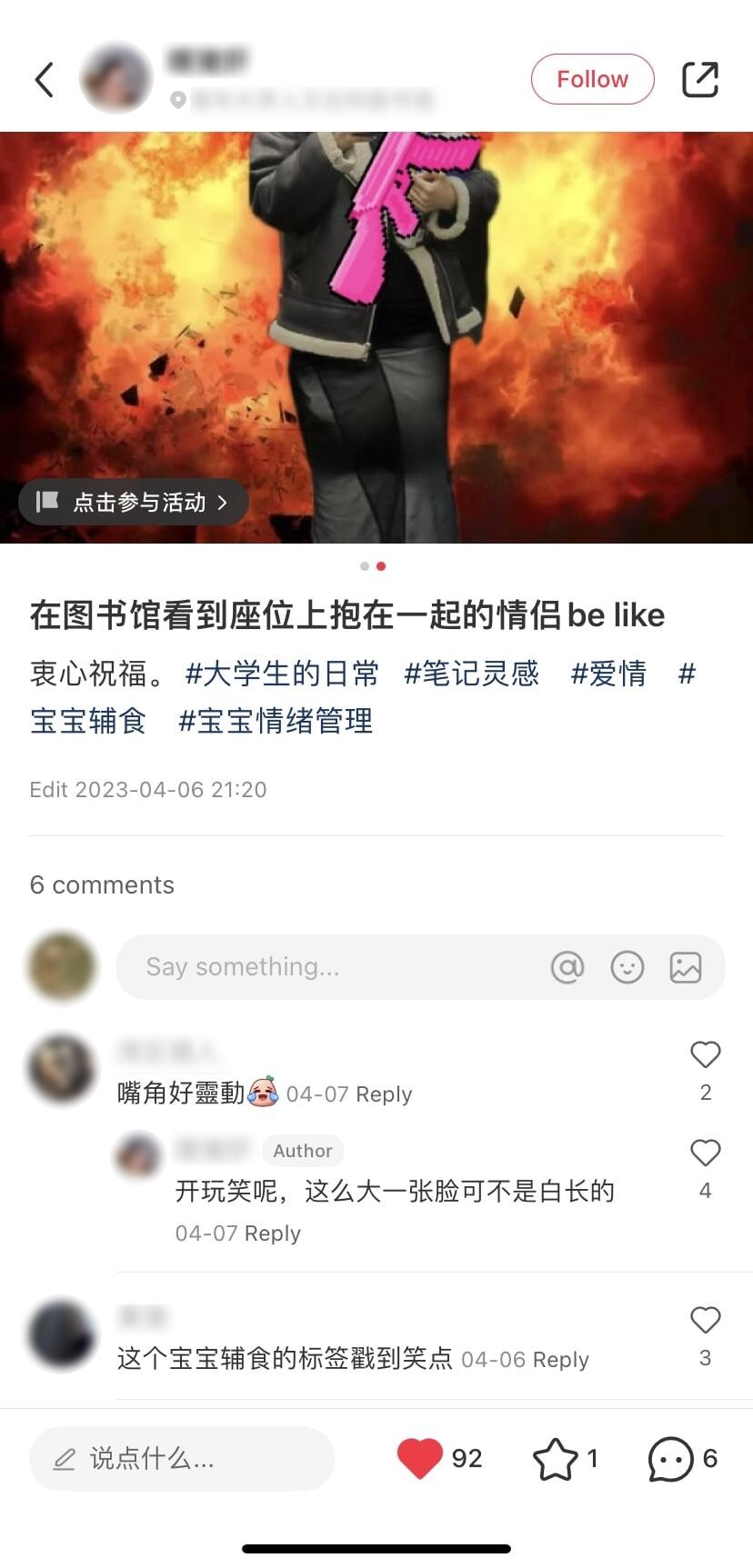}
        \caption*{(2) Post page}
    \end{minipage}
    \caption{Xiaohongshu's primary interfaces include: (1) the Explore page, a two-column feed of recommended posts that users can scroll through and click on for more details; and (2) the Post page, a full-screen display where users can comment, like, collect, and follow the post creator. Functional hashtags appear as clickable blue text within the post content. Screenshots from August 2024.}
    \Description{Two phone screenshots of Xiaohongshu's interface. (1) The Explore page features a small header with a menu icon on the left, three navigation options: ``Follow'', ``Explore'' (selected), and ``Nearby'', and a search icon on the right. Below, there are interest categories: ``For You'' (selected), ``Videos'', ``Like'', ``Fashion'', ``Food'', and a button to expand more categories. The middle of the screen shows a two-column list of recommended posts, with content blurred for privacy. The footer contains icons for Home, Shop, a red ``+'' button to create a post, Messages (with an unread notification), and Me. (2) The Post page includes a header with a back arrow, the poster's avatar and nickname, a follow button, and a share icon. The post occupies the middle section, containing an image at the top, text content with hashtags in blue, and a comment area. All visible text and images are blurred for privacy. The footer displays a text box for comments, along with icons showing the number of likes, saves, and comments.} 
    \label{xhs-screenshot}
\end{figure}

\subsection{Folk Theory and User Adaptive Tactics in Algorithmic Social Media}
In algorithmic social media, content visibility and distribution are often complex and opaque, leading users to form their own folk theories---\textit{``intuitive, informed theories that individuals develop to explain the outcomes effects, or consequences of technologies systems, which guide reactions to and behavior towards said systems''} ~\cite{devito2017algorithms}. The HCI community recognizes folk theory's importance in interpreting users-algorithms dynamics~\cite{Devito2017affordance,devito2018people,eslami2016first,karizat2021algorithmic,devito2017algorithms,Devito2021adaptive, lee2022algorithmic}.
\citet{devito2018people} outlines the folk theory formation process in social media as a cycle of information foraging, sense-making, and theory formation, which guides behavior and refines the theory. 
Similarly, ~\citet{lee2022algorithmic} describes how people understand their identities and view others through personalized algorithmic systems as algorithmic crystals. It uses the metaphor of a crystal to depict how algorithms reflect and refract users' identities, which are dynamic and multifaceted. 
Additionally, other study shows users often believe algorithms suppress content tied to marginalized social identities ~\cite{karizat2021algorithmic}. From past research, users often leverage their folk theories post-hoc to make sense of their experiences, reacting to the platform and guessing the effectiveness of their resistant actions ~\cite{devito2018people}. Users have developed various strategies with the intent to control algorithms: hashtags, coded languages, (dis)like, reporting, etc~\cite{burrell2019users, kim2021trkic, steen2023you}. 
Complementing folk theories, ~\citet{domestication} discuss domestication theory, that users proactively tailor social media to meet their goals through iterative processes of appropriation, objectification, incorporation, and conversion. 

Compared to making sensing of algorithms, content visibility and audience management usually evolve with high-level complexity, which in turn provides users with more flexibility and adaptability ~\cite{cotter2019playing}. 
~\citet{Devito2017affordance} identify feedback directness and audience role as key factors influencing user behavior. To manage their audience, users employ various audience-reach and audience-limiting tools, like hashtags, audience lists, and privacy settings.
~\citet{burrell_when_2019} note that algorithm control strategies are often collectively developed and acted, leading to considerable social utility. For instance, ~\citet{bishop2019managing} found beauty vloggers productively manage their visibility on YouTube through ``algorithm gossip,'' the communal and socially informed strategies to maintain visibility and financial consistency. 

The exploration of folk theories and adaptive tactics in algorithmic social media reveals a nuanced landscape where users continuously adapt their presence and visibility. This dynamic interaction highlights the need for a deeper understanding of how users proactively audit, challenge, and redefine algorithmic boundaries.

\section{The Case of Baby Supplemental Food Hashtag Usage on Xiaohongshu}
\everypar{\looseness=-1}

Established in 2013, Xiaohongshu was initially designed for women consumers focused on cross-border e-commerce. The primary content included beauty product reviews and shopping tips. Up to September 2023, nearly 70\% Xiaohongshu users were women, and 50\% were born after 1995 ~\cite{xiaohongshu_gender,xiaohongshu_age}. Over the last couple of years, it has transformed into a comprehensive social media platform, attracting a broader spectrum of users. Now, Xiaohongshu supports sharing diverse topics including hobbies, entertainment, and daily life, via text, images, short videos, and live-streaming. 

The recommendation algorithm is central to content creation and consumption on Xiaohongshu. Figure \ref{xhs-screenshot} shows Xiaohongshu's user interfaces, where the ``Explore'' page is the default scrolling page in the center of tab options. The page is functioned by personalized recommendation algorithms. By continuously scrolling through the Explore page, users can see an endless stream of posts recommended based on their interests, mixed with a small portion of posts from the users they follow. The recommendations may be influenced by the interests selected in users' settings, the feedback they provide on the recommendations, and their past behaviors (such as searches, clicks, likes, comments, and saved posts), which imply their preferences. The recommendation algorithm references user interests in various, unpredictable ways. This uncertainty also means that content creators may not know who will see their posts or what factors influence the algorithm to distribute their content.

Hashtags are commonly used to define post topics and help target relevant audiences. However, due to the complexity and dynamics of the algorithm, hashtags are not always effective or precise enough for users to control who sees their posts. This limitation can result in unexpected audience reach and potential harassment from unintended viewers ~\cite{klassen2021more}. Xiaohongshu allows individual users to block or report others, or adjust their privacy settings.  
Privacy settings allow users to make posts public, friends-only, private, or visible/invisible to selected accounts. But, these settings cannot prevent certain groups of unknown users who are not welcome seeing the posts in their feeds.  

Recently, hashtags like \#BSF have been used to combat unexpected audience reach on Xiaohongshu. Women users widely used \#BSF in posts unrelated to baby food. This usage, far from the hashtag's literal meaning, sparked curiosity and discussions among users about why it appears in unrelated posts. 
This case of \#BSF hashtag usage on Xiaohongshu provides a unique lens to explore hashtag affordance, gender dynamics, and how users manipulate recommendation algorithms to create safe spaces.

\section{Methods}

We take a mixed-methods approach, including quantitative analysis of Xiaohongshu posts and interviews with Xiaohongshu users. This integration allows us to cross-validate findings and provides a deeper understanding of \#BSF re-appropriation on Xiaohongshu.

\begin{figure*}[t]
    \centering
    \includegraphics[width=1\linewidth]{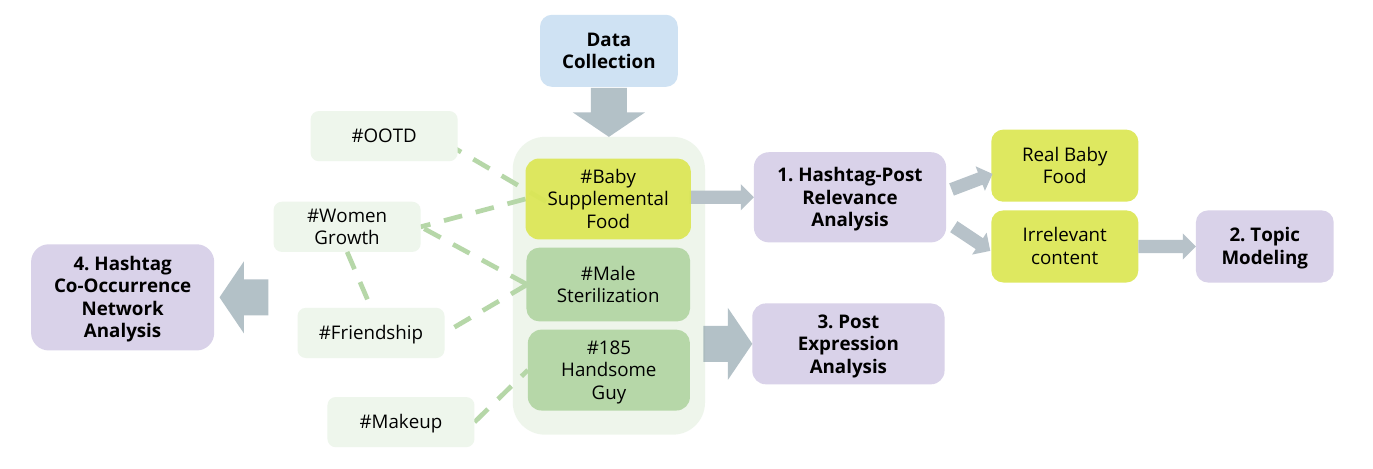}
    \caption{The four steps in the process of quantitative data collection and analysis. 1) \textbf{Hashtag-Post Relevance Analysis} for \#BSF posts classifies content as relevant or irrelevant. 2) \textbf{Topic Modeling} further identifies topics within those irrelevant posts. 3) \textbf{Post Expression Analysis} examines the linguistic features of posts with \#BSF and derived hashtags. 4) \textbf{Hashtag Co-Occurrence Network Analysis} explores the relationships between \#BSF, derived hashtags, and their co-occurred hashtags.}
\label{fig:quant_framework}
    \Description{A flow chart illustrating the quantitative analysis procedure. The chart consists of several rectangular elements connected by arrows to show the workflow. It begins with a rectangle labeled ``Data Collection,'' which leads to a group of hashtags including \#Baby Supplemental Food, \#Male Sterilization, and \#185 Handsome Guy, representing that we collected posts with these hashtags. The \#Baby Supplemental Food leads to ``Hashtag-Post Relevance Analysis'' and then branches into two labeled categories: ``Real Baby Foo'' and ``Irrelevant Content,'' with ``Irrelevant content'' further connected to ``Topic Modeling.'' Additionally, all collected hashtags lead to ``Post Expression Analysis.'' Finally, all collected hashtags and their co-occurred hashtags, including \#OOTD, \#Women Growth, \#Friendship, and \#Makeup, are connected by dashed lines and further lead to ``Hashtag Co-Occurrence Network Analysis''.}
\end{figure*}

\subsection{Positionality}
\label{sec: positionality}
The co-first and the last author are Chinese women who are active users of Xiaohongshu. The other co-authors include two Chinese-speaking male researchers and a White woman with a background in East Asian languages and cultures. The co-first and last authors were responsible for data analysis, while the other authors contributed to discussing the analysis results and providing mentorship. We acknowledge that the authors' backgrounds and experiences might influence the interpretation of the analysis. To minimize potential biases, reduce subjectivity, and deepen the core findings, the authors engaged in iterative discussions to incorporate the perspectives of men and non-Asians.

\subsection{Quantitative Analysis of Hashtag Usage Patterns}
The quantitative data collection and analysis processes are illustrated in Figure~\ref{fig:quant_framework}. We collected posts containing \#BSF, its derived hashtags, and a few general baseline hashtags. For \#BSF, we analyzed hashtag-post relevance and used topic modeling to explore hashtag re-appropriation. Next, we examined post expression to compare the different sentiments and attitudes across hashtags. Finally, we conducted a hashtag co-occurrence network analysis to identify patterns of hashtag use and differentiation among user groups.

\subsubsection{Data Collection}
\label{sec:data collection}
We collected posts with the original \#BSF hashtag and its derived hashtags from Xiaohongshu from February to April 2024 through web scraping. For each post, we collected its title, content, and hashtags.
As \#BSF gained popularity, its usage became more complex. Additional hashtags with similar purposes, such as \begin{CJK*}{UTF8}{bsmi}\#寶寶輔食\end{CJK*} (BSF in Traditional Chinese, abbreviated as BSF(TC)) and \begin{CJK*}{UTF8}{gbsn}\#男性结扎\end{CJK*} (Male Sterilization, abbreviated as MS), also emerged ~\cite{sohu2024}.
Then, we identified three popular hashtags derived from \#BSF through empirical observations of user discussions and co-use with \#BSF on Xiaohongshu, and collected posts using these hashtags. The derived hashtags include \#BSF(TC), \#MS, and \begin{CJK*}{UTF8}{gbsn}\#185大帅哥\end{CJK*} (185 Handsome Guy, abbreviated as HG).

We also constructed a baseline dataset that represents the general post content on Xiaohongshu. We selected three hashtags with high exposures and active users---\begin{CJK*}{UTF8}{gbsn}\#美食\end{CJK*} (Cuisine), \begin{CJK*}{UTF8}{gbsn}\#我的日常\end{CJK*} (My Daily), and \begin{CJK*}{UTF8}{gbsn}\#不懂就问有问必答\end{CJK*} (Ask if you don't know, answer if someone asks)---and collected their posts. These three hashtags correspond to three popular types of posts we observed on Xiaohongshu: interest, daily sharing, and Q\&A. These posts were then compared with those containing \#BSF and its derived hashtags in the following analysis. Table \ref{tab:data} provides a summary of all selected hashtags and collected posts.

\begin{table*}
    \centering
    \begin{tabular}{p{0.08\textwidth}p{0.4\textwidth}p{0.2\textwidth}p{0.16\textwidth}}
        \toprule
        \textbf{Group} & \textbf{Hashtag} & \# \textbf{Total Views (billion)} & \textbf{\# Posts Collected} \\ \midrule
        Original & \begin{CJK*}{UTF8}{gbsn}宝宝辅食\end{CJK*} Baby Supplemental Food, BSF & 29.75 & 3,311 \\ \midrule
        Derived & \begin{CJK*}{UTF8}{bsmi}寶寶輔食\end{CJK*} BSF in Traditional Chinese, BSF(TC) & 0.083 & 324 \\
        & \begin{CJK*}{UTF8}{gbsn}185大帅哥\end{CJK*} 185 Handsome Guy, HG & 2.88 & 613  \\
        & \begin{CJK*}{UTF8}{gbsn}男性结扎\end{CJK*} Male Sterilization, MS & 0.2 & 338 \\ \midrule
        Baseline & \begin{CJK*}{UTF8}{gbsn}美食\end{CJK*} Cuisine  & 58.04 & 450 \\
        & \begin{CJK*}{UTF8}{gbsn}我的日常\end{CJK*} My Daily & 81.27 & 455 \\
        & \begin{CJK*}{UTF8}{gbsn}不懂就问有问必答\end{CJK*} Ask if you don't know, answer if someone asks &  37.38 & 309 \\ 
        \bottomrule
    \end{tabular}
    \caption{Summary of collected hashtags and posts from Xiaohongshu. The hashtags are divided into three groups: \textbf{1) Original}: the hashtag \#BSF, which is the primary focus of the study; \textbf{2) Derived}: hashtags with similar usage to \#BSF; and \textbf{3) Baseline}: popular general hashtags on Xiaohongshu for comparison. Total views are the numbers displayed on the hashtag board in Xiaohongshu as of August 19, 2024.
}
    \label{tab:data}
\end{table*}

\subsubsection{Hashtag-Post Relevance Analysis}
\label{sec: relevance analysis}
Hashtags can either relate to the post content or not. For the \#BSF hashtag, 
we define relevant content as posts that are directly about baby food, such as a recipe for supplemental porridge. In contrast, posts of parenting tips, like table manners, as well as posts of women's selfies, would be classified as irrelevant, even though they may relate to broader parenting or lifestyle topics. 
A prevalence of irrelevant content indicates that the hashtag has deviated from its original purpose, a phenomenon often observed in cases of audience selection. 

To quantitatively identify the prevalence of irrelevant content with the posts containing the \#BSF hashtag, we conducted a hashtag-post relevance analysis. We categorized each post into one of two categories: 1) \textit{Relevant}: the post is about baby food, and 2) \textit{Irrelevant}: the post is about anything other than baby food. Specifically, we adopted GPT-4o-mini ~\cite{achiam2023gpt} as the classifier through few-shot prompting ~\cite{gpt3}, where we provide the model with a few examples within the following prompt: \textit{``Below is the title of a social media post. Classify the post into one of the two categories: A. It is about baby food. B. Everything else. Examples for category A: [A list of 16 example post titles]. Return with only `A' or `B'. POST TITLE: [Post title to be classified].''}

To validate the model’s accuracy, we first tested it on a manually annotated test set consisting of 100 posts, including 32 Relevant posts and 68 Irrelevant posts\footnote{Two authors manually annotated the 100 posts, achieving a Cohen's kappa score of 0.977, indicating an almost perfect agreement ~\cite{cohen1960coefficient}.}. We intentionally included a higher proportion of Relevant posts to address the issue of sample imbalance when calculating accuracy. The model achieved an accuracy of 94\% on the test dataset. We also tested the condition of classifying using both post titles and content---i.e., replacing all titles in the prompt with the titles and their content---which resulted in lower accuracy (75\%). This may be due to the challenges of the longer input text. Therefore, we chose to classify using only post titles. Out of the 3,311 posts, we classified 502 as Relevant and 2,809 as Irrelevant. 

\subsubsection{Post Topic Modeling}
\label{sec: topic analysis}
To further investigate the derived use of the \#BSF hashtag, we conducted topic modeling upon the 2,809 posts that were classified as Irrelevant. We used BERTopic\footnote{BERT: Bidirectional Encoder Representations from Transformers, a language model. } ~\cite{grootendorst2022bertopic}, a pipeline that uses transformers and the Class-based Term Frequency-Inverse Document Frequency (c-TF-IDF) to create dense clusters with interpretable topics. Specifically, we utilized SentenceBERT ~\cite{reimers-2020-multilingual-sentence-bert}, UMAP\footnote{UMAP: Uniform Manifold Approximation and Projection.} ~\cite{mcinnes2018umap}, and HDBScan\footnote{HDBScan: Hierarchical Density-Based Spatial Clustering of Applications with Noise.} ~\cite{mcinnes2017hdbscan} for text embedding, dimension reduction, and clustering. We iterated the program through grid search and manual checks, resulting in 28 clusters\footnote{Details including model hyperparameters and full clustering results can be found in Appendix \ref{topic-modeling}.}. We named each cluster based on the top keywords selected by c-TF-IDF, the label generated by GPT-3.5-turbo ~\cite{achiam2023gpt}, and the content of the posts within the cluster. Finally, we visualized the topic clusters using two-dimension embeddings reduced by UMAP with BERTopic.

\subsubsection{Post Expression Analysis}
\label{sec: liwc analysis}
The linguistic expressions of posts can reflect the purposes, emotions, and interaction targets of the original post creators. Posts with \#BSF and its derived hashtags showed unique features in language use compared to the general content on Xiaohongshu. Quantifying these features helps us understand the characteristics of these hashtags on a larger scale and cross-validate our qualitative observations.

We adopted a lexicon-based approach by using the Chinese linguistic inquiry and word count (LIWC) dictionary ~\cite{huang2012development}, a specialized lexicon designed for analyzing Chinese text. The dictionary includes vocabulary covering grammatical, psychological, and content categories. From these, we selected 12 categories pertinent to our research: \textit{positive emotion}, \textit{negative emotion}, \textit{social}, \textit{family}, \textit{female}, \textit{male}, \textit{body}, \textit{sexual}, \textit{ingest}, \textit{power}, \textit{anger}, and \textit{swear}. Explanations and example words of these categories can be found in Appendix~\ref{liwc}. We established two groups of datasets: a baseline dataset combining posts from the three general hashtags, and four experimental datasets consisting of posts with the \#BSF hashtag and its three derived hashtags. To compare the datasets, we applied Tukey's Honest Significant Difference (HSD) test and conducted pairwise comparisons of means across all five datasets for each linguistic category. This test is particularly appropriate as it controls the family-wise error rate across multiple comparisons, minimizing the probability of making Type I errors while comparing the means of multiple datasets simultaneously.

\subsubsection{Hashtag Co-Occurrence Network Analysis}
\label{sec: network analysis}
To further understand the relationship between the \#BSF hashtag, its derived hashtags, and other related hashtags, we conducted hashtag co-occurrence network analysis. Hashtags appearing in the same post are considered co-occurring. The co-occurrence hashtag network shows how users employ hashtags to associate different topics within their posts.

Each post on Xiaohongshu can contain up to 10 hashtags. Hashtags were identified as any phrase prefixed with the ``\#'' symbol and suffixed with a space. We extracted all hashtags from posts containing the \#BSF and derived hashtags, then generated a co-occurrence matrix where each cell $(i,j)$ represents the number of posts in which hashtag $i$ and $j$ co-occurred. Using this co-occurrence matrix, we constructed undirected, weighted networks. In the networks, nodes represent unique hashtags, and edges represent the co-occurrence between two hashtags, with edge weights indicating the frequency of co-occurrence. Higher weights indicate more frequent co-occurrence. By adjusting the threshold of the edge weights, we visualized the co-occurrence networks to highlight the most frequent co-occurrences while reducing the noise from hashtags that appear only occasionally.

\subsection{Qualitative Interview Study with Xiaohongshu Users}

We conducted 24 semi-structured online interviews from June to July 2024 to understand Xiaohongshu users' perspectives on the usage of the \#BSF and its derived hashtags. Our study was approved by our institution's Institutional Review Board.

\subsubsection{Recruitment}
\label{sec:interview recruitment}

We recruited 24 participants through Xiaohongshu (n = 18), Mastodon (n = 3), personal connections (n = 2), and Instagram (n = 1). Participants must be over 18 years old, and users of Xiaohongshu, as the focus of this study is on the platform's use and engagement with specific hashtags. Participants should have either used or encountered the studied hashtags on Xiaohongshu. These hashtags include but are not limited to \#BSF. Once interest was expressed, individuals would receive a detailed overview of the study, including its purpose, what participation involves, the expected duration of their involvement, any potential risks and benefits, and their rights as participants. After confirming the consent, participants would engage in a one-hour semi-structured online interview. Each participant received a cash compensation of 50 Chinese Yuan (equivalent to \$6.92 USD) after finishing the interview. 

To include comprehensive perspectives, we recruited diverse participants, including 17 female (70.83\%) and seven male users (29.17\%) following the users' gender ratio on Xiaohongshu. Among them, 4 participants have children; 5 participants disclosed their LGBTQ+ identity; 10 participants have experience in either operating marketing content for companies or operating personal content as self-identified independent content creators. We refer to participants’ sexual orientations based on their self-identification, including only those who voluntarily disclosed this information during interviews. Additionally, we obtained their explicit consent for its inclusion in the paper. More details about the participants are presented in Table \ref{tab:Participants}.

\subsubsection{Interview Procedure}
\label{sec: interview}
Before formal interviews, we conducted two pilot interviews with a female and a male Xiaohongshu user to refine the interview questions. After the pilots, the co-first authors and the last author reviewed the process together. We then removed some confusing or less relevant interview questions and added questions that emerged during the conversation but were not included in the original interview protocol. 
For instance, we added ``What do you think of those hashtags' usage'' to understand users' perception of the re-appropriated hashtags, and removed ``How do you think algorithms affect content presentation and user interaction under these hashtags?'' as it is not a question users would usually think about. The pilot interviews are not part of the final dataset. 

In the formal interview, we began with learning about participants' regular usage of Xiaohongshu to build rapport (e.g., ``What kind of content do you usually browse or post on Xiaohongshu?''). We then spent about half an hour discussing their experiences and thoughts on the \#BSF hashtag (e.g., ``Do you remember the first time you used or saw the `Baby Supplemental Food' hashtag? What was your initial
impression or experience?''). Afterward, we asked if the participants had encountered audience selection hashtags other than \#BSF. If they could not recall any specific hashtags, we would elicit the discussion using derived hashtags listed in Table \ref{tab:data}. In the remaining half hour, we asked about their experiences and thoughts on gender bias (e.g., ``Have you experienced gender biased comments on Xiaohongshu?''), and their views and strategies for online safety. The full interview protocol and its translation can be found in Appendix \ref{Protocol}.
All interviews were conducted in Mandarin Chinese and audio recorded via Zoom or Tencent Meeting based on participants' preferences.

\subsubsection{Transcript Analysis}
\label{sec: transcript analysis}
Interview recordings were transcribed into Chinese and de-identified before analysis. We conducted a thematic analysis ~\cite{braun2006using} to uncover themes in the data.
Initially, the co-first authors and the last author coded one interview independently and discussed the codes to create a preliminary codebook, which contained the initial codes that were agreed upon by all authors. The last author then wrote a memo for each interview to summarize the key points. Two authors further independently coded three additional interviews (P05, P07, and P17). 
We picked these three interviews because each of them represents a unique user group. 
In our pool of participants, there are three main profile characteristics: whether they have children, whether they identify as LGBTQ+, and whether they have content operation experience. 
P05 is a female professional content operator with deeper insights into content operation strategy than other operators.
P07 is a mother of two children who shared richer insights into childcare than other parent participants. 
P17 is a gay man who provided more unique LGBTQ+ experiences on Xiaohongshu than other LGBTQ+ participants.
Then, the two authors discussed together to synthesize an intermediate codebook and arrived at 69 categories and 197 codes.
The 20 remaining interviews were split evenly between the two co-first authors for independent coding. Throughout the process, the two authors iteratively discussed and refined the codebook together. By the end, the codebook contained a total of 584 codes. 

To finalize the themes, we then analyzed all codes and cross-referenced with the memos to identify the salient patterns in the interview transcripts. 
Given the nuances and complexity of the re-appropriation of hashtag usage, we consolidated 13 categories and grouped them into the following three high-level themes: behaviors of hashtag re-appropriation, motivations of hashtag re-appropriation, and reactions to hashtag re-appropriation after it became popular. Two categories did not fit into any of these themes, so we classified them under others. Table \ref{tab:theme_development} provides a summary of the high-level themes, categories, category descriptions, and example codes.

\section{Findings}
\everypar{\looseness=-1}

Given the mixed-method analysis, our findings reveal hashtag re-appropriation's evolution process, underlying motivations, diverse user groups' reactions, as well as the comparative affordance of hashtags in audience management and content relevance.

\subsection{Process of Hashtag Re-Appropriation: Blocking, Attracting, and Evolving}
\label{sec: process}
In this section, we explore the practice and development of hashtag re-appropriation from the post analysis and interviews. We find most posts using the re-appropriated \#BSF are unrelated to the hashtag's literal meaning but cover various topics (e.g., makeup, pets), demonstrating its effect in blocking male users. The interviews revealed that women users experimented with, discussed, and promoted this blocking strategy to other users. As \#BSF gained popularity, other users began using it to drive traffic. In response, women users developed additional hashtags (e.g., \#HG, \#MS) evolving from \#BSF, which maintained a similar blocking effect.

\subsubsection{Relevance Analysis and Topic Modeling in the Posts}
\label{sec: relevance_topic_analysis}
We first illustrate how the \#BSF hashtag was re-appropriated through quantitative analysis results. In the hashtag-post relevance analysis, out of a total of 3,311 posts with the \#BSF hashtag, 502 posts were categorized as Relevant to baby food (15.2\%), while the remaining 2,809 posts were categorized as Irrelevant (84.8\%). This contrasts with the \#Cuisine hashtag, where all 450 posts collected were identified as food-related. This suggests that the \#BSF hashtag has been significantly re-appropriated from its original literal meaning and usage.

\begin{figure}[t]
    \centering
    \includegraphics[width=1\linewidth]{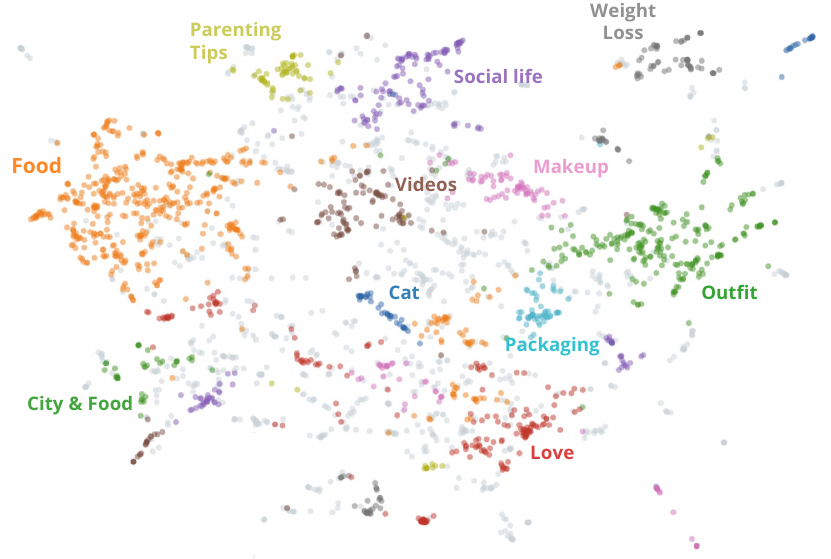}
    \caption{Topic clusters among \#BSF posts that were irrelevant to baby food. Each point represents one post. In descending order, the ten largest clusters are 1) Food, 2) Outfit, 3) Love, 4) Social Life, 5) Videos, 6) Makeup, 7) Weight Loss, 8) Parenting Tips, 9) Packaging, and 10) Cats.}
    \label{fig:topic_scatter}
    \Description{A scatter plot with points of various colors, where each color represents a different cluster. Each cluster is labeled with a short description of its theme, such as ``Food'' or ``Outfit.'' Grey dots represent unclassified posts.}
\end{figure}

Figure~\ref{fig:topic_scatter} visualizes the top ten topic clusters of the 2,809 posts classified as Irrelevant. The largest cluster is still \textit{Food}, though not specifically baby supplements. Most of these posts are related to food photos, recipes, and restaurant reviews. Three of the top ten clusters are about women's appearance: \textit{Outfit}, \textit{Makeup}, and \textit{Weight Loss}, which ranked 2nd, 6th, and 7th, respectively. The \textit{Love} topic ranked the third, with many of the posts talking about real-life romantic relationships, or fantasies---romance novels, movies, and games. A significant portion of posts in the \textit{Love} cluster are from lesbians, who often use the \#Le hashtag at the same time. Next is \textit{Social Life}, which contains posts where women seek/share advice with each other on everything from careers to interpersonal relationships. This is followed by the \textit{Video} topic, containing mostly vlog posts. The top ten topics also include \textit{Cats}, where users share daily photos and updates about cats and exchange adoption information.

The topic modeling illustrates that posts with \#BSF cover diverse topics, indicating the popularity of \#BSF among women users on Xiaohongshu. As it became a trending meme, some users started using \#BSF to gain higher exposure. This included not only regular users who found it funny or trendy but also users with specific intentions, such as advertisers or thirst trap posters. While advertisements do not form a specific topic cluster, they exist in multiple clusters. For example, in the \textit{Video} cluster, there are ads for cameras and handcraft. In the \textit{Parenting tips} cluster, there are soft ads for baby vitamins and body lotions. In the \textit{Packaging} cluster, fans sell celebrity merchandise and albums to others. Thirst trap posts were also present in the data we collected but were not identified in the topic modeling, likely because most of these posts contained explicit photos with cryptic or irrelevant titles and content, making them difficult to identify as a distinct topic.

\begin{figure*}[t]
    \centering\includegraphics[width=0.71\linewidth]{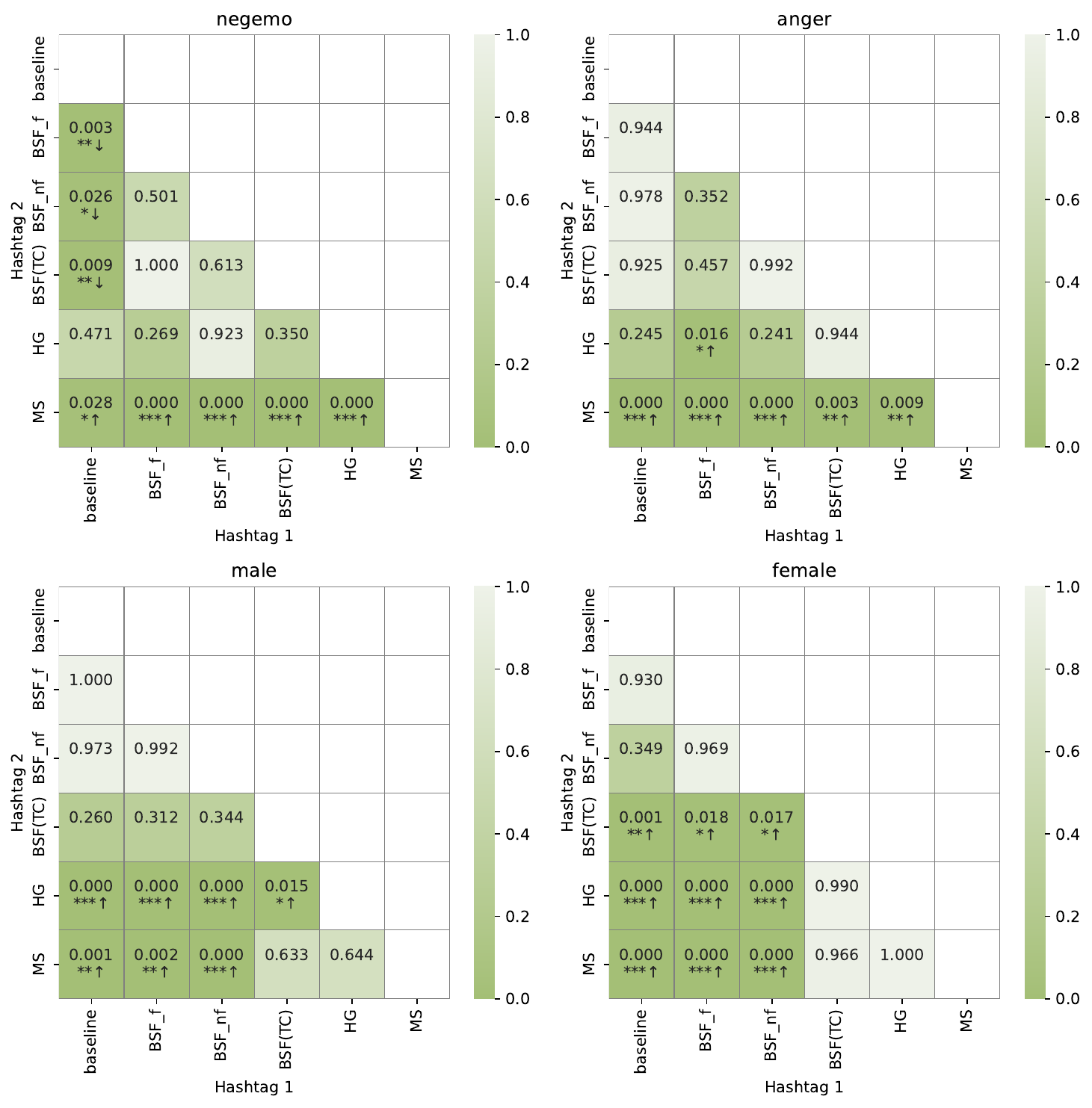}
    \caption{Results of post expression analysis using the Tukey HSD test. Six datasets are compared: baseline, \#BSF\_f (\#BSF posts relevant to baby food), BSF\_nf (\#BSF posts irrelevant to baby food), \#BSF(TC), \#HG, and \#MS. Each matrix corresponds to a specific LIWC category: negative emotions (`negemo'), anger (`anger'), male referent (`male`), and female referent (`female'). Each cell in the matrix indicates whether the expressions of the two datasets in the given LIWC category differ significantly. A deeper color represents a lower p-value, indicating a larger difference. Significant results are marked with `*', `**', and `***', indicating p < 0.05, p < 0.01, and p < 0.001, respectively. An upward arrow ($\uparrow$) suggests that the value for the Hashtag 2 dataset is higher than for the Hashtag 1 dataset.}
    \label{fig:enter-label}
    \Description{Four matrices showing the Tukey HSD test results, each corresponding to one LIWC category. Both the x-axis and y-axis include the six datasets being compared, and the cell shows the result of the pairwise comparison.}
\end{figure*}

\subsubsection{Strategy Learning and Sharing Among Women Users}
\label{sec: learning}
From our interviews, we learned that women users began re-appropriating the \#BSF hashtag as early as 2022. Many women users initially encountered the re-appropriation usage on their ``Explore'' pages, where the hashtag was used in unrelated posts. Driven by curiosity, some users asked about it in the comments, searched for more information, or discussed it with other Xiaohongshu users. They discovered that the hashtag was re-appropriated to prevent male users from seeing posts because the women content creators thought men would not be interested in childcare topics. After learning this, many female users picked up this strategy and began using the hashtag in their own posts.

Positive feedback in audience control using \#BSF validated women users' expectations of the strategy, encouraging them to continue using it. P05 and P08 both mentioned that they received less number of views and also fewer messages from men, which convinced them that this tactic worked.  
But negative feedback also discouraged some women who were trying it for the first time. P12 (Lesbian) shared her experience of posting a photo of herself in a bikini during an island vacation using \#BSF. She regarded Xiaohongshu as a safe space for discussing women's topics and hoped the algorithm would recommend her content to women users. However, after initially receiving a few comments from women complimenting her photo, the posts soon had more than 20,000 views, primarily from men, who either followed her or left harassing comments, such as \textit{``My girlfriend of three years hasn't even let me see this much.''}, which forced her to delete her photos in the end.
\begin{quote}
    I initially approached it (using \#BSF to block male audience) with a trial mindset, but now I have quite a negative view of this hashtag. I really wanted it to recommend my post to a specific group, but this topic just doesn’t work well. I feel like Xiaohongshu could be a space for discussing women’s issues, but in reality, it isn’t. ---P12, lesbian woman.
\end{quote}
The disappointment of some female users did not stop the rapid spread of this hashtag re-appropriation strategy. Users who were optimistic about the strategy actively engaged in sharing, discussing, and co-creating re-appropriation tips with other peers. They believed \#BSF has been effective in audience control, and brainstormed more alternatives. 
\begin{quote}
    I saw others using it, and we (women users) discussed that it seemed effective. We also talked about how men on the platform rarely post about their families, as they tend to focus on themselves—how great they are or how they direct the world. They never think about their home or family. So we found that aside from baby food and family-related content, 95\% of men are absent from Xiaohongshu, with the remaining 5\% being gay men. ---P05, woman, content operator.    
\end{quote}
Further, women even formed private chat groups on Xiaohongshu to discuss hashtag re-appropriation strategies and experiences (P05).

\subsubsection{Evolution of Hashtag Re-Appropriation}
\label{sec:evolution}


The re-appropriation of the \#BSF hashtag underwent various stages, evolving into diverse forms of hashtag usage over time. Insights from our participants’ experiences and accounts helped construct a relatively comprehensive understanding of this evolution process. Several participants (P03, P08, P13, and P19) noted that they were initially aware of the use of \#BSF as a strategy to block male users. 
\begin{quote}
    The first time I saw the `baby supplemental food' hashtag, I didn’t really understand it. It was on a dancing video where the creator said in the comments, `Men, stop harassing me.' Then someone replied, `If you don’t want men to see your posts, just add the `baby supplemental food' hashtag. That way, most men won’t find it.' That’s how I learned about the hashtag. ---P03, 23 years old, women.
\end{quote}

As \#BSF gained popularity, more participants became aware of this tactic; at the same time, participants (P07, P10, P17, and P24) observed that the usage of \#BSF had become more complex. Male users also started using it; some users used it to drive traffic; and even thirst trap content started appearing with \#BSF. At this point, the hashtag was found to be less effective in blocking male users. 
\begin{quote}
    Now there are people intentionally using the `baby supplemental food' hashtag in thirst trap. I think they’re not doing it to avoid unwanted attention but because the hashtag has become so popular. They’re deliberately using it to attract attention. Before, when it wasn’t widely known, they wouldn’t have used it like this. ---P10, lesbian woman.
\end{quote}

In response, users began creating derived hashtags based on the same assumption of what topics men might not be interested in. These derived hashtags were \#BSF(TC), \#HG, \#MS, etc. Some participants viewed these derived hashtags as potential replacements for \#BSF in audience blocking. Some participants even believed that combining \#BSF with all the derived hashtags could result in a more effective strategy for blocking male users. However, others (P06, P09, and P10) expressed concerns about certain derived hashtags, such as \#MS, citing their aggressive nature. For instance, P06 found \#MS to be offensive and felt uncomfortable using the hashtag. She appreciated \#BSF for its mild tone and appealing name, believing it to be more widely accepted and effective. Meanwhile, traffic-driven content creators also shifted to using other trending hashtags after \#BSF's popularity decreased.




Our post-expression analysis shows that posts with \#BSF or its derived or baseline hashtags each have significantly different patterns of sentiment and gender discussion.
Figure~\ref{fig:enter-label} illustrates the Tukey HSD test results for post expression with \#BSF and its derived hashtags. Despite the fact that \#BSF has become a public anti-male signal, some female users still consider it a mild expression. Female users with stronger anti-male attitudes chose to use \#MS to express their perspectives further. From post-expression analysis with LIWC, posts with \#BSF or \#BSF(TC) have significantly lower negative emotions than baseline Xiaohongshu data ($p < 0.05$), while posts with \#MS have significantly higher negative emotions ($p < 0.05$) and anger ($p < 0.01$) than all other hashtags. 
In addition, the frequency of both female and male references was significantly higher in posts with \#MS or \#HG compared to baseline and posts with \#BSF ($p < 0.01$), whereas only female references were significantly higher for posts with \#BSF(TC) ($p < 0.01$). This is not only because \#MS and \#HG are literally related to men, but also because the re-appropriation of these two hashtags is often associated with gender- or relationship-related topics. Full results of the post-expression analysis can be found in Appendix~\ref{fig:full_liwc}. 
Participants commonly perceive \#MS itself is more aggressive than \#BSF which could make it more powerful while also more vulnerable. 
\begin{quote}
    When the `baby supplemental food' hashtag was no longer effective, I switched to `male sterilization' after seeing others use it. My posts got fewer views, and the hashtag wasn’t widely used. But, one day, my account was banned. I suspect it was tied to the `male sterilization' hashtag—likely because men reported me. When I asked why, Xiaohongshu said it violated community guidelines. Thinking about it, none of my other posts seemed report-worthy, but if some men reported me for using the `male sterilization' hashtag, it likely worked. ---P08, 27 years old, woman.
\end{quote}


\begin{figure*}[t]
    \centering\includegraphics[width=0.9\linewidth]{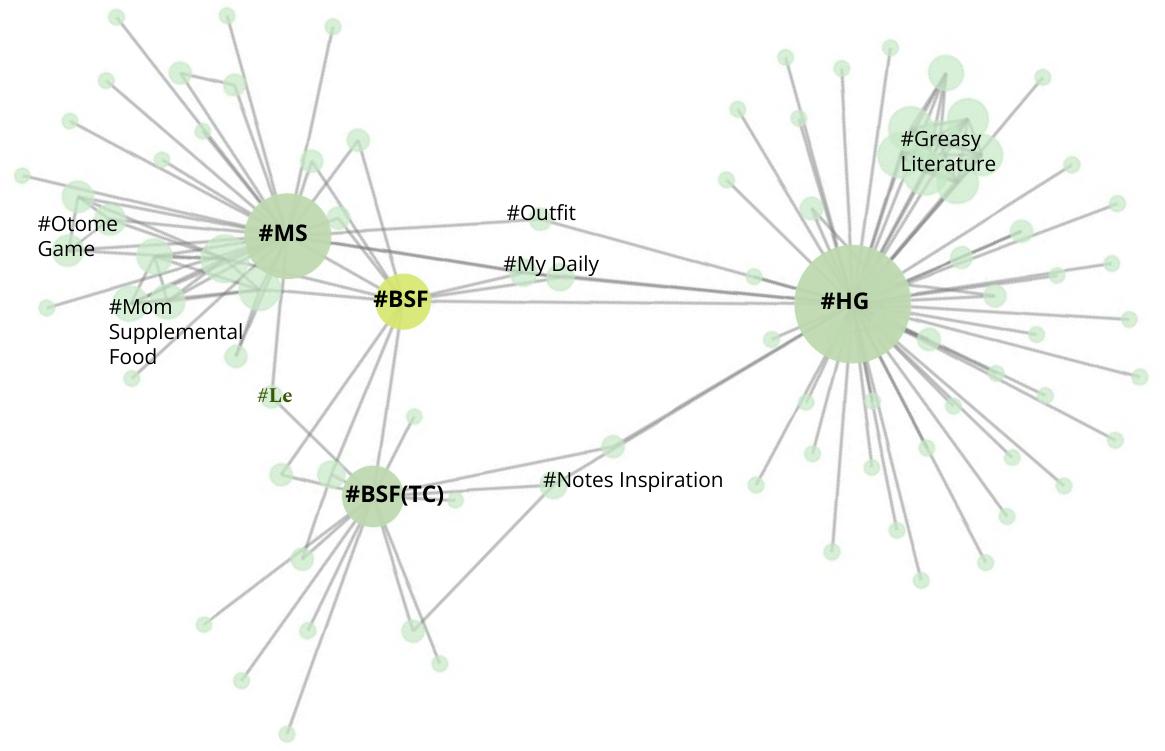}
    \caption{Co-Occurrence Network of Derived Hashtags. Each node represents a hashtag, and each edge indicates that two hashtags co-occurred in a post. Node size reflects the number of hashtags that occurred. \#BSF is positioned at the center of the network, serving as the primary node, and is closely associated with three derived hashtags: \#HG, \#MS, and \#BSF(TC). Each of these forms a distinct co-occurring hashtag cluster.} 
    \label{fig:network}
    \Description{A network diagram consists of circular nodes connected by lines. The central node, \#BSF, is linked to three distinct nodes: \#HG, \#MS, and \#BSF(TC). Each of these nodes is further connected to smaller nodes representing co-occurring hashtags. For example, \#MS connects to \#Otome Game and \#Mom Supplemental Food, while \#HG connects to \#Greasy Literature. \#BSF(TC) and \#HG are linked by the hashtag \#Notes Inspiration.}
\end{figure*}

Further, our co-occurrence network analysis for \#BSF and its derived hashtags depicts a more granular usage pattern of derived hashtags. 
As shown in Figure~\ref{fig:network}, \#BSF, positioned at the core of this network, serves as a primary node, closely associated with each derived hashtag. Each derived hashtag has distinct co-occurring hashtag clusters. For instance, \#HG has a co-occurring hashtag cluster about ``\begin{CJK*}{UTF8}{gbsn}油腻文学\end{CJK*} (Greasy Literature)'', a trendy term in the online community used by people (mostly women) to make fun of men's greasy flirts. \#MS has two main co-occurring hashtag clusters: one focuses on ``\begin{CJK*}{UTF8}{gbsn}乙女游戏\end{CJK*} (Otome Game)'', a genre of female-targeted romance games, and the other on ``\begin{CJK*}{UTF8}{gbsn}妈妈辅食\end{CJK*} (Mom Supplemental Food)''.
Otome Game players enjoy engaging in romantic experiences with virtual male characters within games. They use \#Otome Game to share their game experiences with other peer women players on Xiaohongshu while using \#MS to avoid unwanted attention from male users who usually stigmatize women players. 
\#Mom Supplemental Food is also a re-appropriated hashtag, echoing \#BSF. Notably, posts using \#MS and \#Mom Supplemental Food often feature thirst-trap-like photos. 
Additionally, \#BSF(TC) shows a closer connection with \#MS than with \#HG. \#Le (Lesbian) connects \#BSF(TC) with \#MS, while \#Notes Inspiration connects \#BSF(TC) with \#HG. The hashtag co-occurrence network illustrates how various user groups favor different derived hashtags to achieve their audience control goals and align with their nuanced interests. 
\begin{quote}
    I came across it (\#BSF) while scrolling. It was pretty obvious—there was this girl, really pretty, who clearly identified as a lesbian. I often scroll through similar content where lesbians joke around and share it with friends, so it made sense it was shown to me. In her post, she used the `baby supplemental food' hashtag, and the comments said it’s an effective way to avoid men. Sure enough, it got recommended to me. ---P14, lesbian woman, content operator.
\end{quote}

\begin{figure*}[t]
    \centering \includegraphics[width=0.9\linewidth]{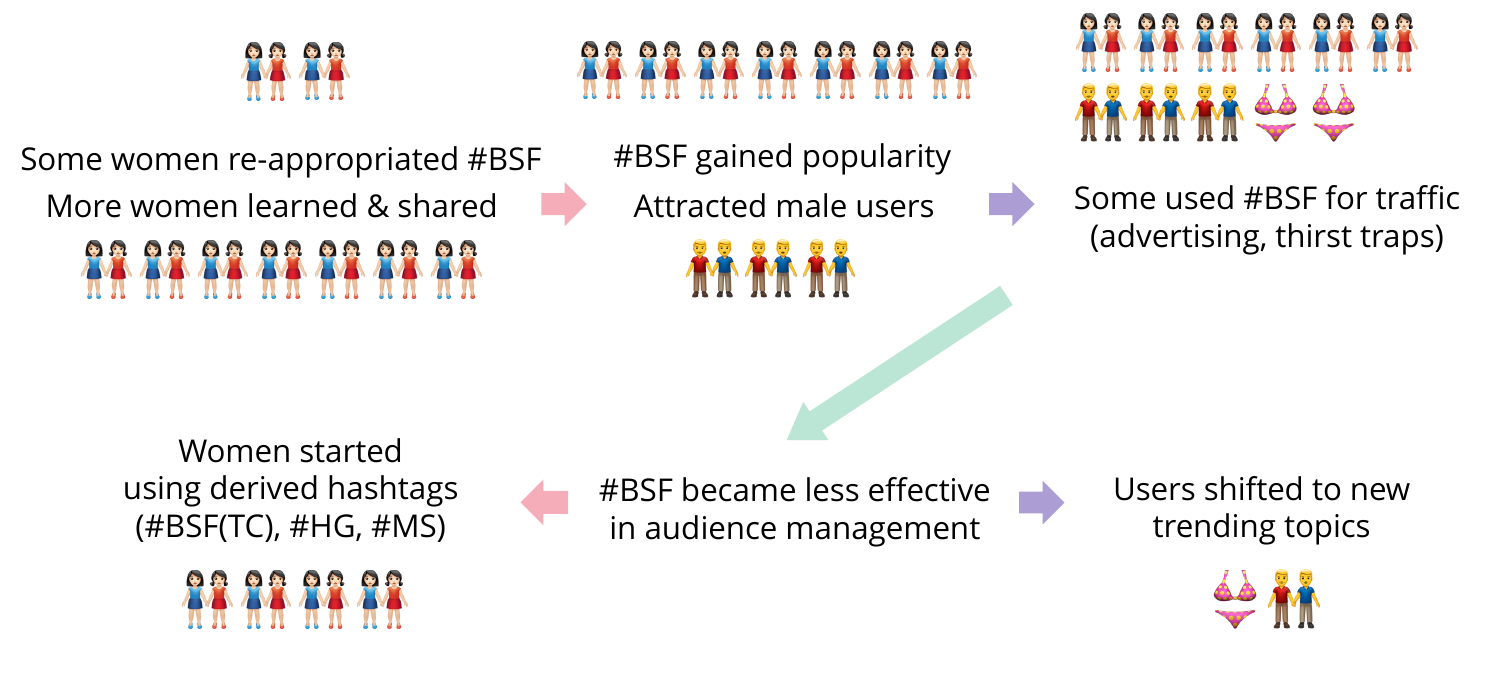}
    \caption{Evolution of the BSF case on Xiaohongshu. Initially, women began re-appropriating \#BSF to protect their posts from male users, which encouraged more women to learn and share. As the hashtag gained popularity, it attracted significant traffic, including male users. Some began using \#BSF for traffic-driven purposes, such as advertising and thirst traps, making it less effective for managing audiences and losing its novelty. As a result, many users shifted to new trending topics. However, women with a persisting need for audience control started adopting derived hashtags like \#BSF(TC), \#HG, and \#MS.
}
    \label{fig:evolve}
    \Description{A flow diagram containing text, emojis, and arrows illustrating the evolution of the \#BSF hashtag. Emojis represent different users and posts: women for female users, men for male users, and bikini emojis for thirst trap posts. The diagram shows the progression of the hashtag's use and shifts in audience behavior over time.}
\end{figure*}

Overall, the \#BSF hashtag re-appropriation went through stages of blocking certain audiences, attracting various audiences given its popularity, and evolving into other derived hashtags. The entire process is illustrated in Figure~\ref{fig:evolve}.

\subsection{Motivations of Hashtag Re-Appropriation}
Analyzing the usage of \#BSF and its derived hashtags paved the way for understanding the underlying reasons behind users' participation in this re-appropriation. Our interviews reveal three primary challenges users faced on Xiaohongshu that led to hashtag re-appropriation: 1) a shifting user base on the platform, 2) persistent gender-based harassment, and 3) the ineffectiveness of existing safety features. 

\subsubsection{Platform Strategy Shift: Introduction of Male Users}
Xiaohongshu started as a women-oriented community, which attracted many loyal women members over the last few years. However, as it expanded its business scope, Xiaohongshu intentionally brought more male users to sign up and engage with the platform. Several women participants mentioned a disruptive incident where Xiaohongshu's advertising department posted pictures of women users to a men-oriented platform, Hupu\footnote{Hupu (Tiger Dab): a Chinese online forum focusing on sports events and men's interests in life. \url{https://www.hupu.com}.}, with a slogan like ``Many pretty women in Xiaohongshu, not spending a penny, free to see!'' Some women participants expressed their disappointment with the platform's use of sexually suggestive promotional tactics (P08, P14, P16). P08 further expressed her frustration with the subsequent influx of male bloggers, who not only joined in large numbers but also consistently received higher traffic on their posts. This signals a shift in the platform’s focus away from its female-centric origins:  
\begin{quote}
    I first realized the platform just cared about money than creating a user-friendly space when it redirected female users’ content to Hupu to attract male users. It's really disappointing. Soon, more male bloggers appeared and got incredible traffic on their posts, especially gay men doing makeup. ---P08, 27 years old, woman.
\end{quote}

Since its inception, Xiaohongshu has been regarded as a space to share women-related narratives, such as romantic relationships, family challenges, and career development. As P09 and P16 commented, Xiaohongshu used to be a place for girls to listen to and support each other. However, with the introduction of male users, gender-sensitive content was exposed to them, leading to many unexpected conflicts and instances of inter-gender harassment. The hashtag re-appropriation action reflected the efforts of women users to preserve this original ``women's space'' for their information-sharing and support needs, while simultaneously resisting the platform’s shift toward a more gender-balanced business model. Given the prevalence of inter-gender harassment reported by participants, we provide a detailed discussion of it in the following section.

\subsubsection{Persistent Gender-Based Harassment}
Notably, all women participants mentioned that they had received and experienced annoying comments, private messages, harassment, or doxxing from men online. These ranged from mansplaining to judging women's body shape, belittling women's professions, and sexual harassment, etc. For example, P05 and P16 mentioned that when they shared their weight loss process, they experienced men's attacks on their body image, such as mocking them for being as big as a tank. P16 (19 years old, woman) also received harassing comments from men such as \textit{``I can take you to do exercise, but on the bed.''} 

Despite that women were the major user group on Xiaohongshu, they often encountered unexpected harms that made them vulnerable. P10 described how the algorithm's unpredictable behavior exposed her personal photos to a wider audience. Then male users passed judgment or unsolicited criticism. This experience led to significant privacy and safety concerns, ultimately prompting her to delete posts and refrain from sharing:
\begin{quote}
    I used to carefully select a few photos and share them on Xiaohongshu. But I feel like sometimes Xiaohongshu’s algorithm would inexplicably push your daily posts to a lot of people. It gives you the feeling of people (male users) pointing fingers at you in the comment section. When that happened, I deleted the post or stopped sharing. ---P10, lesbian woman.
\end{quote}
Women users found themselves constantly on guard, anticipating the next instance of harassment. This eroded their trust in both male users and the platform itself. P05's significant achievement of losing weight was belittled by male users, resulting in a deep sense of frustration:
\begin{quote}
    The year before last it took me 8 months to lose over 50 jin ($\approx$ 55 pounds), a very memorable event for a big fat person, but for them (male users), you're still a tank even after you have done. Well, this was the first time I felt, fuck, the introduction of male users to Xiaohongshu is real crap, driving a car down the drain all at once. ---P05, woman, content operator.
\end{quote}

The harassment toward women users also affected male users when they were mistakenly identified as women. For example, P17 (man) posted that he had obtained his teaching license. Since teaching is stereotypically associated with women, P17 was perceived as a female user by another male user, who commented \textit{``What's it worth for? Doesn't end up going home to raise kids?''}
Similarly, P24 (man, content operator), who used a picture of a woman as his avatar, mentioned that he was often flirted by male users. P24 emphasized that these experiences made him realize how rampant gender-related harassment could be and how annoying it was to be harassed when he was perceived as a woman, even though he was not. 

Female users believed that by using \#BSF, they could create a protected environment for sharing personal thoughts and experiences. Women who wanted to avoid male audiences also found this a smart way of mocking the absence of men in childcare, while also sounding gentle, as women naturally feel an affinity for the topic of babies and motherhood. 

\subsubsection{Ineffectiveness of Reporting and Block Features}
Women participants expressed disappointment with the blocking and reporting features on Xiaohongshu. Most of their reports were unsuccessful. When other users post offensive comments to their posts, they might be irritated and respond more emotionally, leading to conflicts. 
When they reported each other, the platform would punish the female users for their more intense language, not the commenters who instigated the dispute. 
Women users perceived that male users knew how to offend female users and avoid being regulated by the platform. Female users believed that Xiaohongshu's official reporting determination criteria were unreasonable; vicious users had a bottom line in mind when commenting, i.e., they avoided using cursing words so would not be regulated by the platform, but they would stimulate the counter-attacking emotions of female content creators, causing the latter to be regulated given their heated defenses. Eventually, deleting the posts, and silence became the remaining power that those women content creators own.

\begin{quote}
    They (vicious male users) not only have experience, but they've probably discussed this. They know where that boundary is and then will do this kind of thing on purpose to get attention, to stir up emotions, and then to get women attacked by the platform or attacked by more people. ---P05, woman, content operator.
\end{quote}
P04 speculated that the platform might only respond to reports when certain keywords were triggered or when the reported user has been flagged multiple times.
\begin{quote}
    For the platforms, I think they take fewer steps. In fact, when I see something making me feel uncomfortable or disgusting, I will report it to the platform, but usually, they are not successful. I think since these things can be sent out, their keywords or something must not have triggered the platform's mechanism, so I think reporting (on Xiaohongshu) is very unsuccessful. ---P04, 28 years old, woman.
\end{quote}
Blocking is an endless effort and not effective enough for Xiaohongshu.
Participants shared that even after blocking, the recommendation algorithm still pushed content to them. Blocking only prevents seeing content on the user's profile page, but there is still a chance that posts from blocked users will appear on the explore page for both parties. Additionally, women users engaged in proactive blocking as a protective measure; when they came across users who spoke unfriendly or when they identified users whose gender was clearly male, they blocked them to prevent harassment and conflicts.
\begin{quote}
    I now simply avoid any conflict with them (vicious male users) because I've come to accept this idea: real power lies in having the right to remain silent and not needing to explain yourself. So, this is my attitude towards many men now --- they go on and on, but I just don’t want to engage anymore. I block them directly. ---P16, woman, content operator.
\end{quote}
Women users have endured enough harassment and the frustration of failed reporting and blocking efforts. As a result, they chose to delete their posts or use the \#BSF strategy to block male audiences.

\subsection{Post-BSF: Different Groups' Reactions to Hashtag Re-Appropriation}
As the \#BSF hashtag gained popularity and spread beyond its original context, it elicited a variety of responses on Xiaohongshu, ranging from its use as an anti-male signal to a tool for driving traffic. 

\subsubsection{Hashtags as Anti-Male Signals}
\label{sec: signals}
Initially, users adopted this hashtag as a secret code to influence the algorithm, avoid men, and gather women. However, once the intention behind the hashtag became widely known, the secret code turned into a public signal. Women users have since used the hashtag to indicate that, even if male users happened to see the post, they were not welcome to join the discussion:
\begin{quote}
    He (male users) will think that this thing you're posting is not directly related to baby supplemental food and that your heavy use of \#BSF is actually a disguised way of using gender to start a war. But as I said, my intention is simple, when male users see \#BSF, please get out, get out of my way. ---P05, woman,  content operator.
\end{quote}
P14 (lesbian woman, marketing on Xiaohongshu) highlighted that while the hashtag had lost its original function, it played a meaningful role in fostering social connections and identification. She explained that users searching for or engaging with \#BSF were using it to recognize and connect with ``like-minded'' individuals. As she put it, the hashtag helped viewers identify ``allies'' as a public marker of shared interests or values: ``I feel that for those genuinely searching for `baby supplemental food' or using the hashtag to identify, `Oh, you’re an ally' or `You’re an interesting person,' it still serves that kind of purpose.''






\subsubsection{Support Hashtag Re-Appropriation with Reservations}
\label{sec: support_w_res}
Some male users used the \#BSF hashtag to counter the way women were using it, arguing that it was inciting gender conflict and occupying the discourse space originally intended for baby food and parenting topics.
However, in our interviews, mothers with children expressed support for women using their own methods to protect their safe spaces (P05, P07). They also mentioned that parents had various ways to find the information they needed, and the re-appropriation of the \#BSF hashtag had not affected their ability to access relevant information. Fathers, while they felt it would be better not to re-appropriate these hashtags, also expressed an understanding of women's intentions to avoid harassment:


\begin{quote}
    I think it’s better to categorize posts accurately; tagging them arbitrarily is not something we should encourage. Perhaps their (hashtag re-appropriation users) intention isn’t necessarily bad---maybe they just want to avoid harassment from some male users. But still, they shouldn’t use this kind of hashtag. They could have… well, I admit there is indeed a lot of harassment. ---P21, man, father of a 7-year-old.
\end{quote}



On the other side, some women users who do not have children also supported the hashtag re-appropriation with reservation. They expressed concerns about taking over a hashtag that was originally intended for parent communication, but they also stated that re-appropriating the hashtag was a last resort for them. Also, while some women have heard that \#BSF became less effective and the derived hashtag \#MS was being used, they were reluctant to adopt this alternative because they didn't want to be aggressive. 
\begin{quote}
    I'm probably less likely to use the \#MS, but I'm willing to use the \#BSF. I feel like it's just a better word, including it's a little bit gentler, but male sterilization is just a little bit more oppositional, and I'm the kind of person who doesn't like conflict. ---P06, 21 years old, woman.
\end{quote}

\subsubsection{Male Incomprehension of Hashtag Re-Appropriation}
When first introduced to the re-appropriated hashtag, users were usually confused. While female users curiously questioned this usage, many male users were confused because they felt offended (P06). Most male users had not experienced the risks that women had experienced in the patriarchal society. Even if sometimes male users were mistaken for women and were attacked by other men, they did not feel hurt because they knew they were not women. 
\begin{quote}
    \textit{Researcher: So like this post on your teacher's certificate, you received this kind of comment, seems women users cheer with you `sister, how great you are' and some of the male users say `So what, you still have to go home to raise kids', what are your feeling of the polarized comments?} \\
    I don’t have much feeling because I’m a man and I don’t have to go home to raise kids. ---P17, gay man.
\end{quote}

Further, male users did not understand why women re-appropriated hashtags to avoid harassment, because, from their perspective, those hashtags were likely to offend some men, and thus can easily provoke harassment. On the other side, male users did not understand why not just use the built-in blocking and reporting features in the system, avoiding extra inter-user conflicts.  
 
\begin{quote}
    I am mainly emphasizing this, maybe the hashtag is not effective, she (his girl friend) felt that I can't understand a woman's mind, she feels that only needs to type very few words, just four words. And it would be better if it could reduce the male harassment of her, so I do understand this kind of behavior. But if say you added hashtags such as \#Impotence, I think it's already attacking. I would directly block it off, and I would not get into an argument with others. ---P20, 25 years old, man.
\end{quote}

\subsubsection{Hashtag Re-Purposing for Driving Traffic}\label{sec: re_purposing}
The fact that \#BSF became more controversial and widely discussed has led to new changes in how \#BSF was used. Given its high visibility, many users began using the hashtag to drive traffic to their posts, without any intention of blocking others, and with no relation to baby food. Additionally, many marketing accounts started using \#BSF to boost the visibility of their products, particularly those targeting women, such as weight loss and beauty products. Some marketing accounts even began using \#BSF as a persona description for a demographic profile of women aged from 20 to 35 with below-average spending power, leveraging the hashtag for targeted advertising, such as study abroad services. Some regular users also used the hashtag as a meme or just to follow the trend. 

Additionally, with the traffic generated by \#BSF, male users were exposed to and interacted with \#BSF posts that were supposed to be protected from them. As a result, female users found the hashtag less useful for audience selection (P06). Content creators also found it less useful in driving traffic, some even observed a decrease in views compared to posts that did not use the hashtag and suspected that this was related to moderation by the platform (P05, P08).

\subsection{Comparing Hashtag Affordance from Audience Management and Content Relevance}
\begin{figure*}[]
    \centering
    \includegraphics[width=0.6\linewidth]{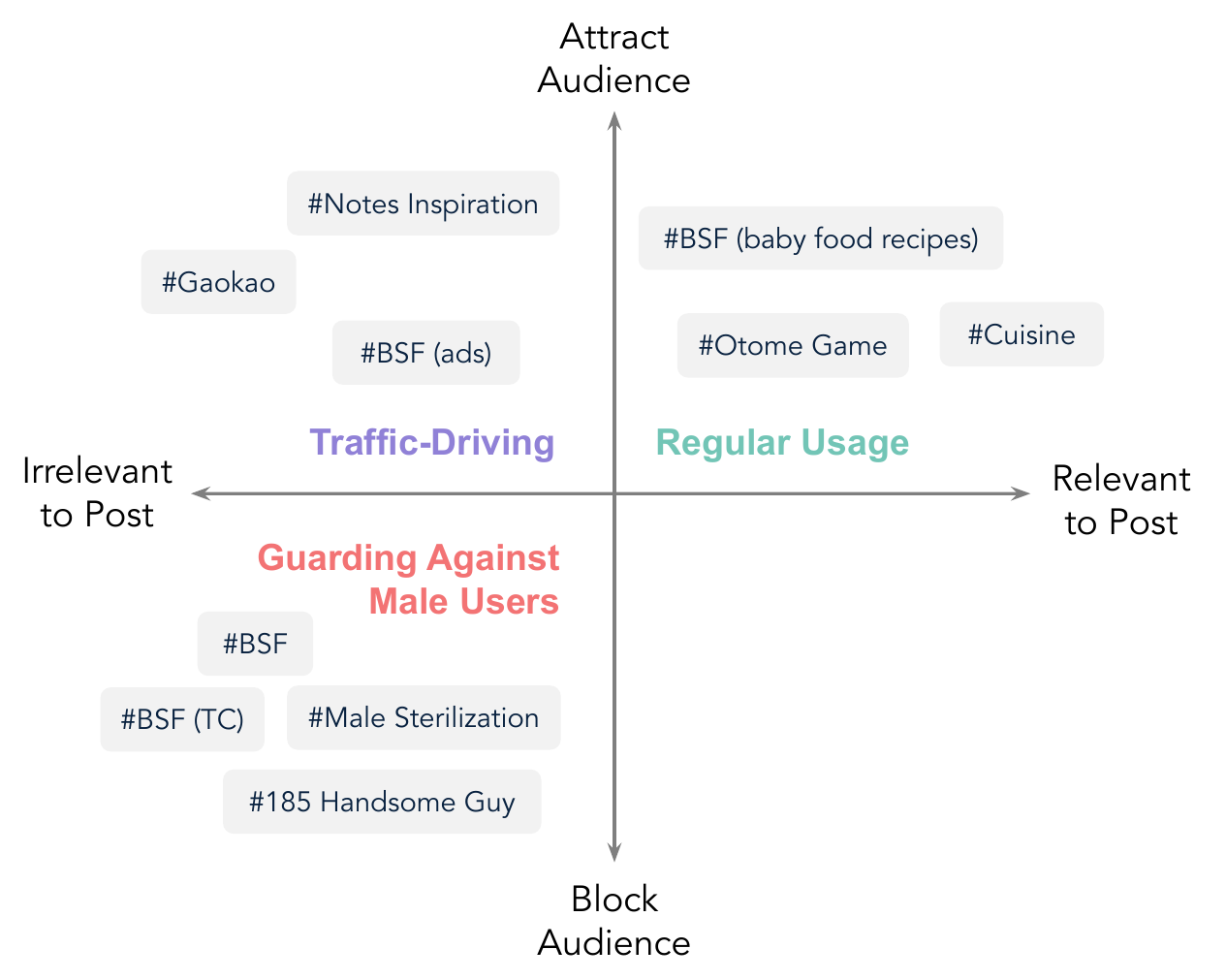}
    \caption{Two-dimensional coordinate illustrating hashtag affordance in terms of audience management (attract vs. block) and content relevance (relevant vs. irrelevant). Based on the quadrants, three types of usage are identified: \textbf{1) Regular Usage}, where hashtags describe the post content and attract the intended audience; \textbf{2) Traffic-Driving}, where hashtags are often irrelevant to the post and are used primarily for attracting traffic; \textbf{3) Guarding Against Male Users}, where \#BSF and derived hashtags are re-appropriated to block audiences. The fourth quadrant (block + relevant) does not correspond to any existing usage and is left blank.}
    \label{fig:tag_usage}
    \Description{A Cartesian coordinate diagram with two axes: the x-axis represents a hashtag's relevance to the post (left: irrelevant, right: relevant), and the y-axis represents whether the hashtag is used to attract or block an audience (top: attract, bottom: block). The diagram is divided into four quadrants, each labeled with its corresponding usage. For example, the top-right quadrant (Attract audience + Relevant to Post) is labeled ``Regular Usage'' and contains hashtag examples shown in gray rectangles.}
\end{figure*}
We synthesized the above findings and compared the roles of hashtags in different use contexts. We found that hashtag affordance could be interpreted from two perspectives: 1) audience management---whether the hashtag is used to attract or block the audience; and 2) content relevance---whether the hashtag is semantically relevant to the post content. This analysis yields three distinct uses of hashtags: regular usage, traffic driving, and guarding against male users, as shown in Figure~\ref{fig:tag_usage}. We discuss the three categories in conjunction with the findings above.

\subsubsection{Regular Usage}
The most common usage is ``Attract \& Relevant'', that is, using hashtags relevant to the post to attract audiences who are interested in the topic. For example, a user can share a homemade dinner using \#\begin{CJK*}{UTF8}{gbsn}美食\end{CJK*} (cuisine) so that the post will be recommended to users who are interested in dinner recipes. Users looking for cuisine in a particular city can search for this hashtag along with the name of the city to find reviews of local restaurants. Specific hashtags relevant to the posts can also be used to attract the audience. For example, \#Otome Game, as mentioned before, is used by players with the same interest to find each other in the hashtag topic board. The original use of \#BSF also falls into this category, attracting parents of infants who wanted to learn how to make baby supplemental food (Section~\ref{sec: relevance_topic_analysis}).

\subsubsection{Traffic Driving}
Some hashtags are not used to describe the content of the post but rather serve as a means to drive traffic. We categorized these hashtags as ``Attract \& Irrelevant.'' Since Xiaoshongshu allows up to ten hashtags for each post, these traffic-driving hashtags typically follow other hashtags that are relevant to the post content. The usage of these hashtags to drive traffic can be initiated either by the platform or by users. The most popular officially-operated hashtag on Xiaohongshu is \#\begin{CJK*}{UTF8}{gbsn}笔记灵感\end{CJK*} (Notes inspiration), which gathers a variety of popular topics in the platform. As shown in the network analysis, it is a common co-occurring hashtag with \#BSF(TC) and \#HG (Section~\ref{sec:evolution}). The hashtag was designed to increase user engagement, in return, the platform promised higher exposure for posts using this hashtag. As a result, many content creators have included this hashtag in their posts. 

Another way to drive traffic is using hashtags trending in real-time on the platform. For example, P19 (man, content operator) mentioned he used \#\begin{CJK*}{UTF8}{gbsn}高考\end{CJK*} (Gaokao, National College Entrance Examination) in his meme posts to drive traffic in June, during the time of the Chinese National College Entrance Exam, which is widely discussed nationwide. 
This approach attracted more people to view these posts, including those who were not originally interested in the topics. 
Some marketing accounts also used these trending hashtags to increase views and reach a wider potential audience (Section~\ref{sec: re_purposing}).

\subsubsection{Guarding Against Male Users}
Both of the categories described above aim to attract audiences, a typical purpose for content creators on social media. However, women users re-appropriated hashtags to block certain audience groups, giving rise to a third category: ``Block \& Irrelevant.'' The re-appropriated \#BSF falls into this category. This practice deviates from hashtags' typical usage---representing post content or driving traffic---and instead leverages the recommendation algorithm to block certain audiences. As \#BSF gained popularity, it gradually lost its effectiveness in blocking male users, leading to the re-appropriation of additional hashtags from their regular usage, including \#BSF(TC), \#HG and \#MS (Section~\ref{sec: signals} and Section~\ref{sec: re_purposing}).

\section{Discussion}
\everypar{\looseness=-1}
Our research uncovered the process of hashtag re-appropriation---blocking, attracting, and evolving. Given ineffective reporting mechanisms and losing trust in Xiaohongshu's women-centered position, women communally re-appropriated hashtags to block unwanted audiences. Different user groups support this re-appropriation with varying levels. 
The section discusses insights from our findings: the shift towards dynamic audience control, hashtag adaptability and limitations, feminism as everyday resistance, user-driven decentralization of power, and limitations and future work.

\subsection{Blocking Affordance: From Static To Dynamic Audience Control}
Our study highlights the ineffectiveness of Xiaohongshu’s blocking and reporting mechanisms, with users frustrated by failed reporting attempts and the platform’s limited blocking features. 
In recommendation-driven platforms, many content filtering features are designed for content consumers to navigate what they want to see in their recommendation feeds, such as clicking ``not interested in such content'' in the recommended items, selecting interested topics in the preference setting, or providing keyword filters in the content preferences setting in TikTok ~\cite{TikTokSupport}. However, fewer features are designed for content creators to control who sees their posts. 
Content creators, in particular, wanted to prevent engagement from unwanted groups while maintaining broad visibility. 
The need to filter unwanted audiences extends across communication platforms. For example, ~\citet{kim2021trkic} found that, in Google Maps Place reviews, Korean speakers crafted text only understandable to fellow speakers to strategically avoid censorship and unwanted attention. 
Similarly, before presenting intended content, TikTok content creators employ creative strategies like starting with topics that may not be interesting for the unwanted audience to prompt them to scroll away---such as women TikTokers starting with discussing menstruation experiences to filter out male audiences ~\cite{TikTokVideo_women}. 
However, content creators' need for robust tools of blocking unwanted audiences remains overlooked.

Our results also reveal gaps in hashtag affordance for balancing audience blocking and content relevance, as shown in Figure~\ref{fig:tag_usage}. While \#BSF offered some audience-blocking capabilities, it failed to maintain content relevance, leading to divergent uses (e.g., traffic driving, thirst trap content). This underscores the need to separate audience control functions from content expression. Static audience controls, such as privacy settings and block/mute features, offer fixed control of a predefined list of audiences but are not adapted to \textit{content-based audience control}. This refers to tailoring audience groups based on the semantic relevance of content to the audience, such as topics they are interested in or they are posting about.  
Hashtags were re-appropriated for content-based audience control given their flexible and rich semantic expression. They enabled users to define and identify unwanted audiences through natural language expression. 

To address the audience control needs, we propose \textit{dynamic audience control.} 
Different from static privacy settings and indirect content moderation strategies, dynamic audience control allows creators to define a content-based selection of audience and directly control audience reach. This dynamic audience control builds a connection between the semantic meaning of content and its audience, capturing evolving content features associated with unwanted audiences rather than targeting specific individual accounts. For example, misogyny can be linked to user accounts that promote or support it. By configuring the keyword “misogyny,” one can create a list of unwanted audiences related to this theme. 
This creates a semi-public environment where the content is visible to some users but hidden from others via content-based filters. 
One design option is dividing hashtags into both attracting and blocking functions, with blocking hashtags integrated into visibility settings. 
Recommendation systems also commonly build on personas for users to deliver explainable recommendations ~\cite{barkan2020explainable}, however, these algorithmic filtering approaches have the limitation on their accuracy. 
Another design option draws inspiration from Twitter Blockbots, enabling users to dynamically filter block-worthy accounts using collectively curated lists~\cite{geiger2016bot}. This approach empowers users to regain their agency when engaging in public discourse.

\subsection{Hashtag Re-Appropriation’s Adaptability and Limitations for User Agency}
The re-appropriation of hashtags like \#BSF demonstrates the adaptability of the hashtag mechanism. Users creatively repurposed these hashtags and derived new hashtags to manage their audience and protect against unwanted attention, demonstrating user agency in navigating algorithmic social platforms. Users adapted hashtags for both blocking and self-expression, transforming a tool meant for topic categorization into one for digital activism.  

Hashtags are widely used for digital activism, where large numbers of posts under the same hashtag deliver a collective message to address power imbalances and drive social change  ~\cite{yang2016narrative}. The narrative form created by these interconnected postings empowers people's agency in digital space. 
For example, \#BlackLivesMatter movement fosters a narrative of racial justice ~\cite{ince2017social}, and \#MeToo movement creates a digital space for survivors to share their stories to combat sexual harassment ~\cite{mueller2021demographic}.  
This narrative agency is ``communal, invented, skillful and protean,'' drawing power from its form, content, and social context, as proposed by Campbell's rhetorical framework ~\cite{campbell2005agency, yang2016narrative}. Similarly, the narrative agency in \#BSF emerged to create safe spaces away from the male gaze. The choice of baby supplemental food reflects women's intimacy with motherhood and care. 
However, the hashtag re-appropriation strategy also has its limitations. As hashtags like \#BSF became popular, they were increasingly co-opted by other users, diluting their original purpose (Section~\ref{sec: re_purposing}). This overexposure led to eventual ineffectiveness, reducing its capacity to provide the safe space that women users attempt to create. 
The derived hashtags, such as \#HG and \#MS then shifted the narrative from avoidant protection to empowerment with a female gaze and mocking patriarchy (Section~\ref{sec: process}). 
The hashtag adaptability and limitations also showed in other online communities, such as pro-eating disorder groups on Instagram, where users circumvented platform moderation by developing increasingly complex lexical variations of hashtags to promote dangerous behavior ~\cite{chancellor2016thyghgapp}. As original hashtags were suppressed, these communities intensified their engagement through other coded terms, often resulting in more toxic and vulnerable content. These cases demonstrate when users' original space is destructed, they become more creative and determined in their tactics, whether for protection, empowerment, or evasion from moderation. The mechanisms behind hashtag re-appropriation illustrate users continue to find ways to express agency within platforms' constraints.

\subsection{Feminism Everyday Resistance}
Our study found many participants did not perceive hashtag re-appropriation as directly linked to feminism activism. However, as Vinthagen and Johansson argue, everyday resistance acts are often not explicitly recognized by actors themselves as resistance, which subtly responds to power dynamics within specific social and cultural contexts~\cite{vinthagen2013everyday}. For instance, Nicaraguan women joking about men may not see their actions as resistance. Subalterns rarely define what they do as resistance since they are invisible or marginalized in public discourse, yet these actions quietly challenge and renegotiate power dynamics. 

Similarly, Xiaohongshu women users' hashtag re-appropriation represents a semi-conscious form of everyday resistance. By re-appropriating hashtags, women users protect themselves from male gaze and harassment. Even those who support women with reservations see it as an opportunity to raise awareness of women's struggles and promote safer spaces (Section~\ref{sec: support_w_res}). 
Everyday resistance illustrates how women users maintain control through small-scale actions, even if they do not consciously frame it as resistance. It is important for researchers to uncover and analyze unrecognized assumptions, power dynamics, discursive structures, and potential changes in different forms of resistance. 
The re-purposing of \#BSF and its derived hashtags reflects the heterogeneity of everyday resistance, as women adaptively respond to harassment and unwanted audiences on algorithmic social platforms (Section~\ref{sec: process}). While not being confrontational in the traditional sense, this resistance continuously challenges power structures on digital platforms on a day-to-day basis. 

Digital Chinese feminism's everyday resistance expands beyond Xiaohongshu. Women have employed various methods to create safer, women-only digital spaces. For example, on Weibo\footnote{Weibo: a Chinese social platform, where users post, share, and comment on various topics. \url{https://weibo.com/}.} and Douban\footnote{Douban: a Chinese social platform that allows users to create and join interest-based groups to discuss cultural, social, and lifestyle topics. \url{https://www.douban.com/}.}, women have established private women-only groups to foster communities away from male-dominated spaces. 
Similarly, Otome games and their gaming communities on social media offered women peer support away from being stigmatized and marginalized in the patriarchal social context  ~\cite{lei2024game}. Our research further illustrates that Otome game players use \#BSF and \#MS to build a female-centered peer community to share gaming experiences on Xiaohongshu (Section~\ref{sec:evolution}). 
Meiyou (beautiful grapefruit)\footnote{Meiyou: a Chinese health management app focusing on women menstruation. \url{https://meiyou.com}.}, initially a menstrual tracking app, has also expanded to a feminine communication platform. Up to July 2024, Meiyou has become the third most popular women's health app globally ~\cite{meiyou_statista2024}. 
While these spaces are also not spared from harassment in the patriarchal society, women have continuously adapted tactics, skillfully empowering their own presentation rather than limiting themselves to unchangeable environmental challenges. 

\subsection{Decentralization on Algorithm-Centered Recommendation Platforms}


The \#BSF hashtag re-appropriation on Xiaohongshu exemplifies a case of bottom-up decentralization on \textit{commons-based} platforms. 
According to ~\citet{zhang2024form}, commons-based platforms share users' content to the entire platform, rather than limiting sharing within networks or distinct spaces (e.g., groups). 
Decentralized governance models in online platforms take various forms. ~\citet{jhaver2023decentralizing}'s middle-level governance model decentralizes power through multi-level governance units in online platforms. Mastodon gives users autonomy to run self-hosted instances, each with its content moderation policies.
The shared block lists on Twitter are a bottom-up example that applies a third-party tool (e.g., Blockbots) to enable decentralization through collectively curating block-worthy users by volunteers~\cite{geiger2016bot}. 
On Xiaohongshu, users can only control their own posts with content visibility centrally governed by the platform’s algorithm. Despite this, female users creatively repurposed hashtags like \#BSF as a tool for audience blocking---an unintended use not supported by the platform nor aided by any third-party tools.
This demonstrated vertical cross-level governance, which influences the platform’s centralized recommendation algorithm in content distribution at the top and the content visibility for end users at the bottom. 
Through this re-appropriation practice, female users re-allocated their power in the algorithm-centered social platform. 
However, the effectiveness of the user-driven strategy was not sustainable as the platform's recommendation algorithm exposed its content to broader audiences. This highlights the tension between the algorithm's control of content distribution and visibility and the decentralized user efforts of managing their safe spaces.

\subsection{Limitations And Future Work}
While our study provides valuable insights into hashtag re-appropriation on Xiaohongshu, several limitations present opportunities for future research. First, we had difficulty recruiting participants using \#BSF for thirst trap content, limiting our understanding of their perspectives. Future studies can explore strategies to balance tensions between exploitation, membership, disclosure, and allyship when conducting research with marginalized populations~\cite{liang2021embracing}.

Another limitation is the challenge of systematically gathering data on a recommendation-driven platform. It is infeasible to determine if the data is biased or fully representative. In addition, our study analyzed hashtag usage from a post level, which provides a snapshot of the phenomenon. A longer-term analysis could reveal the trends, algorithm shifts, user engagement, and platform updates. Future research can also incorporate analysis of other data including comments and likes, which provides a more holistic understanding of the communal dynamics. 

We hope our research inspires future studies to investigate user needs and platform affordance across cultures. For example, ~\citet{ibrahim2024islamically} explored how Muslim women use menstrual tracking applications for both health and religious practices, such as managing obligations during menstruation. This highlights the importance of HCI research working closely with users who experience technology limitations given their intersectional backgrounds, pushing the boundaries of inclusive, user-centered design.

\section{Conclusion}
Recommendation-driven social media offers personalized content but often leads to unwanted interactions. 
This study examined how women users on the recommendation-driven platform Xiaohongshu re-appropriated hashtags like \#BSF to create safe spaces. Using a mixed-methods study, we analyzed hashtag-post relevance, post topics, post expression's linguistic patterns, hashtag co-occurrence relationships, and interviews with 24 diverse Xiaohongshu users.  
Faced with persistent harassment, platform value shift, and ineffective safety features, women re-appropriated hashtags to block unwanted attention and promote self-expression. Over time, the hashtag was co-opted for other purposes, leading women to create alternative hashtags for continued audience control with intense attitudes. The case highlights a form of everyday resistance within digital feminism in China and broader implications for audience control and user self-governance on recommendation-driven platforms. We hope this study inspires further research on how marginalized groups navigate algorithmic governance to maintain safe spaces online.

\begin{acks}
We thank the participants of this study for their contributions. We also acknowledge the reviewers for their valuable feedback and suggestions. Special thanks to Rebecca Jones for providing examples of TikTok posts. This research was supported by funding from the Inequality in America Initiative at Harvard University.
\end{acks}

\bibliographystyle{ACM-Reference-Format}
\bibliography{sample-base,reference-qf}

\appendix
\appendix
\section{Interview Protocol}
\label{Protocol}
Here are the questions that guide our one-hour semi-structured interview. 
\begin{enumerate}
    \item Can you share when you approximately started using Xiaohongshu? What is your usual usage frequency?
    \item What kind of content do you usually browse or post on Xiaohongshu?
    \item Do you follow or interact with real-life friends on Xiaohongshu? Why?
    \item Which other social media platforms do you use?
    \item Do you remember the first time you used or saw the  ``Baby Supplemental Food'' hashtag? What was your initial impression or experience?
    \item What is your opinion on this hashtag? Positive, negative, or other views?
    \item When did you start engaging with these hashtags? Has your attitude towards them changed?
    \item Why do you think the hashtag is used? What kind of users use it?
    \item Do you think it is effective?
    \item Why do you think you came across the content using the ``Baby Supplemental Food'' hashtag? Have you followed parenting content?
    \item Do you interact with posts using ``Baby Supplemental Food''? Why?
    \item What hashtags do you usually use when posting? Why? Are they effective?
    \item Have you used the ``Baby Supplemental Food'' hashtag? Why? Is it effective?
    \item How many posts have you published with the ``Baby Supplemental Food'' hashtag? What were they about? Did you use other hashtags simultaneously?
    \item Have you used similar hashtags like ``male sterilization''? Why? Are they effective? 
    \item Is there a difference in interactions or the number of views between posts with or without the ``Baby Supplemental Food'' hashtag?
    \item Have you encountered anyone questioning the use of such hashtags? 
    \item How has your usage of Xiaohongshu changed before and after engaging with the ``Baby Supplemental Food'' hashtag?
    \item Have you found a community on Xiaohongshu? 
    \item Do you look at the tab of posts from users you follow or the tab of recommended ones? Why?
    \item Have you had trouble finding previously interesting content? How did you solve it?
    \item Do you think users of the ``Baby Supplemental Food'' hashtag form a community? Do you consider yourself a member of this community?
    \item What kind of users participate in and use the ``Baby Supplemental Food'' hashtag?
    \item How is the ``Baby Supplemental Food'' topic community different from other groups on Xiaohongshu?
    \item Have you seen any gender bias-related discussions under the posts using the ``Baby Supplemental Food'' hashtag? Do you remember what the discussion was like?
    \item Are you aware of content related to feminism, independent women, or women's rights? What do you think about it? Do you think these are related to the use of the ``Baby Supplemental Food'' hashtag? Does the case of the ``Baby Supplemental Food'' hashtag contribute to the spread of feminism in society?
    \item What do you think is the gender ratio on Xiaohongshu? What about other platforms? Why?
    \item Do you think the case of the ``Baby Supplemental Food'' hashtag exists on other platforms? Why?
    \item Are there people you prefer not to interact with for a safer online experience? How do you avoid them?
    \item Do you feel safe communicating on Xiaohongshu compared to other platforms?
    \item What aspects of online safety are important to you? What methods do you use to maintain your safety?
    \item If you were to rank the safety of different social media platforms, how would you rank them? Why?
    \item Do you have any suggestions for hashtag and algorithm design?
    \item Is there anything I haven't asked that you would like to add?
\end{enumerate}


\section{Participants Details}
Table \ref{tab:Participants} shows the details of our participants' background information.

\begin{table*}[h]
    \centering
    \begin{tabular}{lllllllll}
    \toprule
       \textbf{P\#}  & \textbf{Gender} & \textbf{Age} & \textbf{Education} & \textbf{\makecell{Have\\Children}} & \textbf{LGBTQ+} & \textbf{\makecell{Content\\Operation}} & \textbf{Location} & \textbf{Recruitment}  \\
    \midrule
       P01  & Female & 25 & Master & $\times$ & $\checkmark$  & $\checkmark$ & U.S. & Personal \\
       P02 & Female & 24 & Master & $\times$ & $\times$ & $\checkmark$ & Hong Kong & Xiaohongshu \\
       P03 & Female & 23 & Master & $\times$ & $\times$ & $\times$ & UK & Xiaohongshu \\
       P04 & Female & 28 & Master & $\times$ & $\times$ & $\times$ & Liaoning & Xiaohongshu \\
       P05 & Female & 32 & Bachelor & $\times$ & $\times$ & $\checkmark$ & Jilin & Xiaohongshu \\
       P06 & Female & 21 & Bachelor & $\times$ & $\times$ & $\times$ & Shandong & Xiaohongshu \\
       P07 & Female & 25 & Secondary School & $\checkmark$ & $\times$  & $\times$ & Shanxi & Xiaohongshu \\
       P08 & Female & 27 & Master & $\times$ & $\times$  & $\times$ & Xian & Xiaohongshu \\
       P09 & Female & 18 & High School & $\times$ & $\times$ & $\times$ & Hunan & Xiaohongshu \\
       P10 & Female & 24 & Bachelor & $\times$ & $\checkmark$ & $\times$ & Hebei& Xiaohongshu \\
       P11 & Female & 32 & Bachelor & $\checkmark$  & $\times$ & $\checkmark$ & Beijing & Personal \\
       P12 & Female & 25 & Master & $\times$ & $\checkmark$ & $\times$ & Shanghai & Xiaohongshu \\
       P13 & Female & 24 & Master & $\times$ & $\times$ & $\checkmark$ & U.S. & Instagram \\
       P14 & Female & 31 & Bachelor & $\times$ & $\checkmark$ & $\checkmark$ & Beijing & Mastodon \\
       P15 & Female & 24 & Bachelor & $\times$ & $\times$ & $\checkmark$ & Sichuan & Mastodon \\
       P16 & Female & 19 & Bachelor & $\times$ & $\times$ & $\checkmark$ & Guangdong & Xiaohongshu \\
       P17 & Male & 27 & Master & $\times$ & $\checkmark$  & $\times$ & Jiangsu & Xiaohongshu \\
       P18 & Female & 21 & Bachelor & $\times$ & $\times$ & $\times$ & Hubei & Mastodon \\
       P19 & Male & 25 & Bachelor & $\times$ & $\times$ & $\checkmark$ & Jiangsu & Xiaohongshu \\
       P20 & Male & 25 & Master & $\times$ & $\times$ & $\times$ & Australia & Xiaohongshu \\
       P21 & Male & 31 & Bachelor & $\checkmark$  & $\times$ & $\times$ & Guangdong & Xiaohongshu \\
       P22 & Male & 21 & Bachelor & $\times$ & $\times$ & $\times$ & Hubei & Xiaohongshu \\
       P23 & Male & 28 & Bachelor & $\checkmark$  & $\times$ & $\times$ & Guangdong & Xiaohongshu \\
       P24 & Male & 24 & Bachelor & $\times$ & $\times$ & $\checkmark$ & Fujian & Xiaohongshu \\
    \bottomrule
    \end{tabular}
    \caption{Summary of participants demographics and background (n = 24). The sample includes 17 females (71\%) and 7 males (29\%), aged 18 to 32. Educational backgrounds consist of 2 with secondary or high school degrees (8\%), 13 with bachelor's degrees (54\%), and 9 with master's degrees (38\%). Four participants have children (17\%), 5 self-identified as LGBTQ+ (21\%), and 10 work in content operations (42\%). Recruitment sources include Xiaohongshu, Mastodon, personal connections, and Instagram.}
    \label{tab:Participants}
\end{table*}

\section{Themes of Interview Transcript Analysis}
Table \ref{tab:theme_development} shows the details of the synthesized themes derived from our interview transcripts.

\begin{table*}[h]
\begin{tabular}{>{\raggedright\arraybackslash}p{0.28\linewidth}|>{\raggedright\arraybackslash}p{0.32\linewidth}|>{\raggedright\arraybackslash}p{0.32\linewidth}}
    \hline
    \textbf{Category} & \textbf{Description} & \textbf{Examples} \\
    \hline
    \multicolumn{3}{l}{\rule{0pt}{4ex} \textbf{Theme 1: Behaviors of Hashtag Re-Appropriation} \rule[-2ex]{0pt}{2ex}} \\
    \hline 
       Hashtag Re-Appropriation in Posts 
       & How users re-appropriate the  hashtags within their posts in various context? & The content using \#BSF, using \#BSF to exclude men, reason of \#BSF being effective \\ 
    \hline
        Strategy Learning and Sharing 
        & How users learn and share strategies for hashtag re-appropriation? & Discuss with other users, create a group chat, look at comments \\
    \hline
        Evolution of Hashtag Re-Appropriation & Track the changing use of the hashtags & First time seen \#BSF, users of \#BSF, derived hashtags \\
    \hline
    \multicolumn{3}{l}{\rule{0pt}{4ex} \textbf{Theme 2: Motivations of Hashtag Re-Appropriation} \rule[-2ex]{0pt}{2ex}} \\
    \hline 
        Persistent Gender-Based Harassment & The ongoing issue of harassment faced by women users & Male harassment topic, the type of attack from men, male users mistaken for female users attacked \\
    \hline
        Platform Strategy Shift: Introduction of Male Users & The platform’s strategic decisions to attract male users & Post Ads to draw men, give male users for traffics, perceived gender ratio \\
    \hline
        Ineffectiveness of Block and Report Features &  The challenges users face when trying to block or report inappropriate behavior & Failed to report, platform is unfair, conditions for successful report \\
    \hline
    \multicolumn{3}{l}{\rule{0pt}{4ex} \textbf{Theme 3: Reactions to Hashtag Re-Appropriation After It Became Popular} \rule[-2ex]{0pt}{2ex}} \\
    \hline 
        Hashtags as Anti-Male Signals & How women users use the anti-male hashtags as signals once they become popular? & Claiming anti-male, combine derived hashtags with \#BSF, women feel provoked \\
    \hline
        Hashtag Re-Purposing for Driving Traffic  & How content creators re-purpose popular hashtags to drive traffic to their posts & Attraction mechanism, platform incentive mechanism, the driving traffic effect of \#BSF \\
    \hline
        Parents Support BSF Re-Appropriation with Reservations & Users who have children support women's self-defense strategies but express concerns & Don't mind anti-male, \#BSF occupy the space, parents find childcare information through specific keywords \\
    \hline
        Male Incomprehension of Anti-Male Hashtag Re-Appropriation  & Male users do not understand why women re-appropriate hashtags for anti-male purposes & Incomprehension, offended, women should not avoid conflicts using \#BSF \\
    \hline
        Perceptions Toward Feminist Discourse  & Users’ views on feminist discourse and the hashtag re-appropriation & \#BSF's association with feminism, ways of promoting feminism, understanding of ``independent women'' \\
    \hline
    \multicolumn{3}{l}{\rule{0pt}{4ex} \textbf{Others that are not belong to the high level themes} \rule[-2ex]{0pt}{2ex}}\\
    \hline
        Platform Features Affordance Dynamics  & How the design and functionality of platform features influence user behavior & Recommendation function, searching function, collection function \\
    \hline
        Experiences on Other Platforms  &  The comparison of users' experiences on Xiaohongshu with other social media platforms & Douyin, Douban, Weibo \\
    \hline
\end{tabular}
\caption{Summary of high-level themes, categories, category descriptions, and example codes.}
\label{tab:theme_development}
\end{table*}

\section{Post Topic Modeling}\label{topic-modeling}

The topic modeling process begins by pre-calculating text embeddings with SentenceBERT, using the \textit{distiluse-base-multilingual-cased-v1} model. The Chinese text is tokenized using the Jieba library. The text is then vectorized using the CountVectorizer, which transforms the tokenized words into a document-term matrix. For dimensionality reduction, a UMAP model is employed with n\_neighbors=7, n\_components=5, and a min\_dist=0.01, using the cosine similarity metric. Finally, clustering is performed with HDBSCAN, which groups similar topics using a minimum cluster size of 15 and the Euclidean distance metric. Random states are set for reproducibility throughout the process. The procedure resulted in a total of 28 topics (Table \ref{tab:full_topic}).

\begin{table*}[h]
\centering
\begin{tabular}{>{\raggedright\arraybackslash}p{0.02\linewidth}|>{\raggedright\arraybackslash}p{0.2\linewidth}|>{\raggedright\arraybackslash}p{0.6\linewidth}|>{\raggedright\arraybackslash}p{0.08\linewidth}}
\hline
\textbf{ID} & \textbf{Topic Name} & \textbf{Top Keywords} & \textbf{\# Posts} \\ \hline
1 & Food & {delicious, eat, make, breakfast, egg, nutrition, child, really, ingredient} & 435 \\ \hline
2 & Outfit & {wear, match, top, summer, clothes, share, early spring, style, skirt} & 251 \\ \hline
3 & Love & {love, like, floating light, mom, circle, wealth, meal, stream, ring} & 130 \\ \hline
4 & Social Life & {will, female lead, say, problem, one, male lead, need, think, together, friend} & 113 \\ \hline
5 & Videos & {Shu, video, captain, shoot, assistant, Dalian, sticker, fries, handmade} & 73 \\ \hline
6 & Makeup & {wife, eyes, makeup, stick, makeup, really, eyeliner, pretty, eyelashes} & 67 \\ \hline
7 & Weight Loss & {jin, weight, bold, thin, number, random writing, write down, click, too much light, drop} & 59 \\ \hline
8 & Parenting Tips & {baby, lotion, body, child, food, water lotion, skin, ingredient, baby food, eat} & 59 \\ \hline
9 & Packaging & {package, card holder, membership, out, membership card, coupon, card, unboxing, no package} & 52 \\ \hline
10 & Cats & {cat, kitten, kitty, cat, food, find, adoption, little kitty, in arms} & 40 \\ \hline
11 & Mom and Baby & {mom, baby, draw, drawing, sister baby, do, drawing style, Valentine's Day, human} & 39 \\ \hline
12 & City \& Food (school) & {Hangzhou, burp, cake, Zhejiang University, Guangzhou, closed down, go to school, blogger, afraid} & 37 \\ \hline
13 & Milk and Eggs & {milk, fried rice, awesome, not relying on, way of eating, very fragrant, boil, boss, change} & 33 \\ \hline
14 & City \& Food (expenses) & {Nanjing, saving money, eat, Suzhou, delicious, lifetime, Tainan, gardener, cheap} & 31 \\ \hline
15 & Dogs & {puppy, dog, doggie, old master, baby, adoption, tolerant, owner, lost} & 31 \\ \hline
16 & Haidilao (hotpot brand) & {lao, Haidilao, little brother, one-stop service, lotus flower, magician, sister, come in, partner} & 28 \\ \hline
17 & The Joy of One Bite & {whole, down, that moment, one bite, joy, instant, flip, s, stunned} & 27 \\ \hline
18 & Unimaginably Delicious & {imagine, dare not, delicious, this must be, few, eat, will, one bite, thin} & 26 \\ \hline
19 & Summer & {summer, look forward to, ahhhh, summer, 16592, stolen, zz, already, sunglasses} & 25 \\ \hline
20 & Weight Loss Plan for Tomorrow & {lose weight, tomorrow, do well, definitely, fitness, first day, girls, Jiangxi, 6025} & 25 \\ \hline
21 & Animal and Children & {one, child, that kid, East Asia, fox, blow, scold, years old, why} & 23 \\ \hline
22 & Looking forward to Summer & {shallow, looking forward, a bit, summer, spring, bah, dusk, spring colors, this color} & 21 \\ \hline
23 & One Bite and my Brain Melted & {brain atrophy, one second, small, picked up, that moment, cerebellum, brain, wretched, brother} & 21 \\ \hline
24 & Weekdays & {Monday, hope, Friday, literature, read, weekend, paper, wear, match} & 20 \\ \hline
25 & City \& Food (south) & {Chengdu, Shenzhen, too ahead, too far ahead, empty stomach, remember, subway station, sinister, Sichuan people} & 19 \\ \hline
26 & The Gold in this Bite & {gold content, understand, one bite, talent, Earth, Sichuan-Chongqing people, one bag, Leshan, die} & 17 \\ \hline
27 & I was Right to Follow Trend on Xiaohongshu & {follow the trend, Xiaohongshu, success, once, Chaoshan, thanks, 2024, tabloid, bangs, article} & 16 \\ \hline
28 & Body Height \& Weight & {158cm130, 163cm140, jin, 155cm120, height, anti-fan, 163, beside, a month} & 16 \\ \hline
\end{tabular}
\caption{Summary of all 28 topic clusters identified from \#BSF posts irrelevant to baby food, sorted in descending order by the number of posts in each cluster. Topic names were first summarized by GPT-3.5-Turbo and then manually refined. Top keywords were computed by c-TF-IDF.}
\label{tab:full_topic}
\end{table*}

\section{Post Expression Analysis}\label{liwc}

\begin{figure*}
    \centering\includegraphics[width=0.95\linewidth]{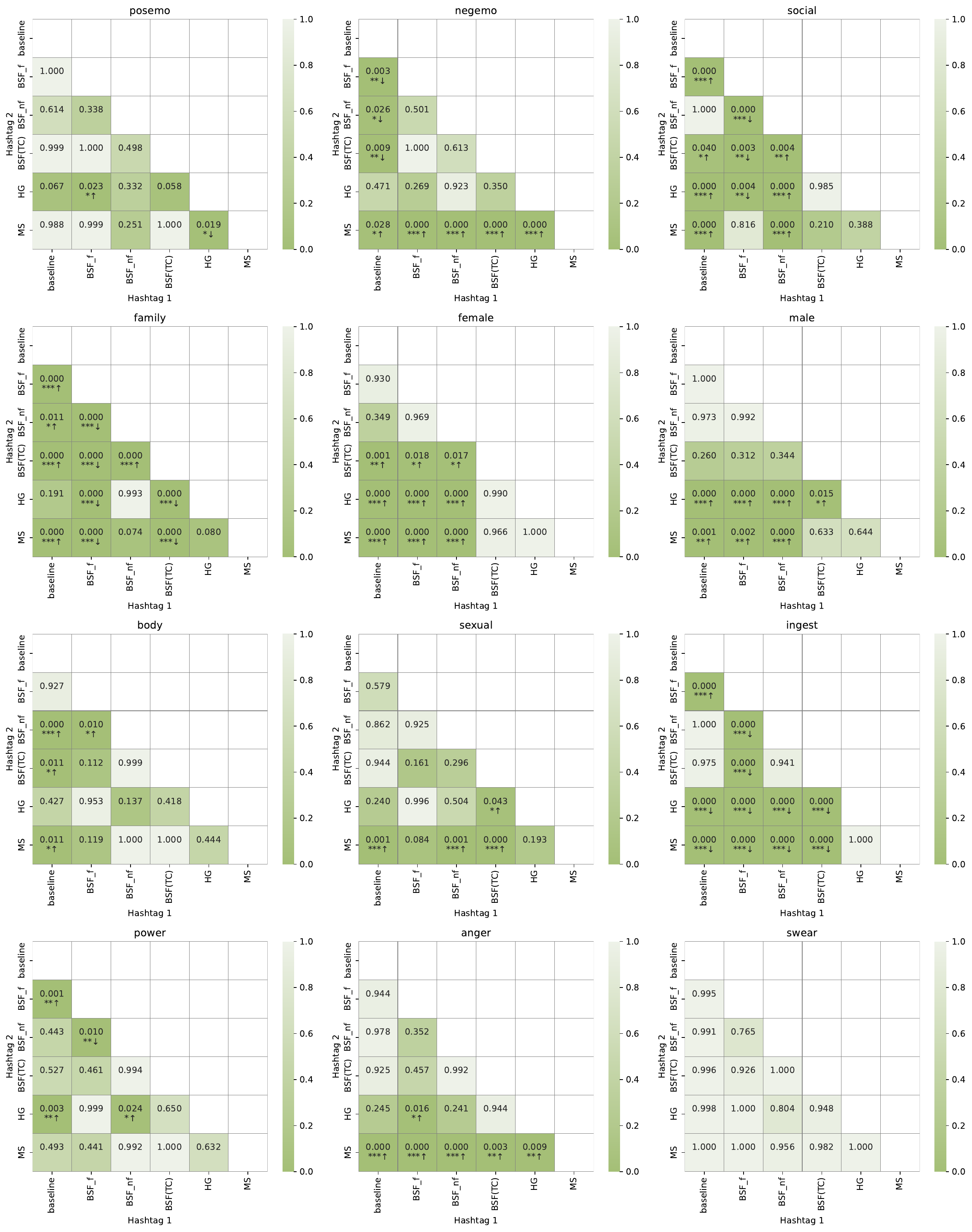}
    \caption{Full results of post-expression analysis using the Tukey HSD test. Each cell in the matrix indicates whether the expressions of the two datasets in the given LIWC category differ significantly. A deeper color represents a lower p-value, indicating a larger difference. Significant results are marked with `*', `**', and `***', indicating p < 0.05, p < 0.01, and p < 0.001, respectively. An upward arrow ($\uparrow$) suggests that the value for the Hashtag 2 dataset is higher than for the Hashtag 1 dataset.}
    \label{fig:full_liwc}
    \Description{12 matrices showing the Tukey HSD test results, each corresponding to one LIWC category. Both the x-axis and y-axis include the six datasets being compared, and the cell shows the result of the pairwise comparison.}
\end{figure*}

We selected 12 LIWC categories that were pertinent to our research. Table \ref{tab:liwc_cat} shows explanations and example words of each category. Figure \ref{fig:full_liwc} shows the full result of a pairwise comparison using the Tukey HSD test. 

\begin{table*}[htbp]
\centering
\begin{tabular}{p{0.1\textwidth}p{0.2\textwidth}p{0.3\textwidth}p{0.3\textwidth}}
\midrule
\textbf{ID} & \textbf{Abbreviation} & \textbf{Category} & \textbf{Examples} \\
\midrule
1 & posemo  & Positive emotion  & love, nice, sweet \\
2 & negemo  & Negative emotion  & hurt, ugly, nasty \\
3 & social  & Social processes  & mate, talk, they \\
4 & family  & Family            & daughter, dad, aunt \\
5 & female  & Female references & girl, her, mom \\
6 & male    & Male references   & boy, his, dad \\
7 & body    & Body              & cheek, hands, spit \\
8 & sexual  & Sexual            & horny, love, incest \\
9 & ingest  & Ingestion         & dish, eat, pizza \\
10 & power  & Power             & superior, bully \\
11 & anger  & Anger             & hate, kill, annoyed \\
12 & swear  & Swear words       & fuck, damn, shit \\
\bottomrule
\end{tabular}
\caption{Selected LIWC categories, abbreviations, and examples.}
\label{tab:liwc_cat}
\end{table*}

\end{CJK*}

\end{document}